\pdfoutput=1
\documentclass[prb,reprint, epsfig,amssymb,bibliography,amsmath,footinbib]{revtex4-2}

\usepackage{blkarray}

\usepackage{amssymb,amsmath,bm,epsfig,graphicx,subfigure,hyperref}

\usepackage[utf8]{inputenc}
\usepackage[vietnamese, english]{babel}
\usepackage{chngcntr}
\usepackage{xcolor}
\usepackage{braket}
\usepackage{physics}
\usepackage{sidecap}
\usepackage{lipsum}
\usepackage{amsfonts}
\begin{document}

\title{Mirror-protected Majorana zero modes in  $f$-wave multilayer graphene superconductors}
\author{\foreignlanguage{vietnamese}{Võ Tiến Phong}$^{1,2,3}$}
\email{vophong@magnet.fsu.edu}
\thanks{These authors contributed equally to this work. }
\author{H\'ector Sainz-Cruz$^{4}$}
\email{hector.sainz@imdea.org}
\thanks{These authors contributed equally to this work. }
\author{Eugene J. Mele$^{1}$}
\email{mele@physics.upenn.edu}
\author{Francisco Guinea$^{4,5}$}
\email{paco.guinea@imdea.org}
\affiliation{$^{1}$Department of Physics and Astronomy, University of Pennsylvania, Philadelphia, PA 19104, U.S.A.}
\affiliation{$^{2}$Department of Physics, Florida State University, Tallahassee, FL 32306, U.S.A.}
\affiliation{$^{3}$ National High Magnetic Field Laboratory, Tallahassee, FL 32310, U.S.A.}
\affiliation{$^{4}$IMDEA Nanoscience, Faraday 9, 28049 Madrid, Spain}
\affiliation{$^{5}$Donostia International Physics Center, Paseo Manuel de Lardiz\'abal 4, 20018 San Sebastián, Spain}
\date{\today}

\begin{abstract}
Inspired by recent experimental discoveries of superconductivity in multilayer graphene, we study models of $f$-wave superconductivity on the honeycomb lattice with arbitrary numbers of layers. For odd numbers of layers, these systems are topologically nontrivial, characterized by a mirror-projected winding number $\nu_\pm = \pm 1$. Along each mirror-preserving edge in armchair nanoribbons, there are two protected Majorana zero modes. These modes are present even if the sample is finite in both directions, such as in rectangular and hexagonal flakes. Crucially, zero modes can also be confined to vortex cores. Finally, we apply these models to twisted bilayer and trilayer systems, which also feature boundary-projected and vortex-confined zero modes. Since vortices are experimentally accessible by local scanning probes, our study suggests that superconducting multilayer graphene systems are promising platforms to create and manipulate Majorana zero modes. 
\end{abstract}

\maketitle

Due to their non-Abelian braiding statistics and  immunity to quantum decoherence, Majorana zero modes (MZMs) are highly sought after as building blocks for topological quantum computation \cite{Ivanov2001Non, kitaev2003fault, Nayak2008Non}. They are believed to exist as excitations of the $\nu = 5/2$ fractional quantum Hall effect \cite{moore1991nonabelions,Read2000Paired}, at the ends of a spinless $p$-wave superconducting chain \cite{kitaev2001}, or in the vortex cores of spin-triplet $p_x+ip_y$ superconductors \cite{Stern2004Geometric, Das2006Proposal, Kraus2009Majorana}. However, these platforms are challenging to realize experimentally because a full understanding of the $\nu = 5/2$ state remains elusive while spin-triplet superconductors are scarce in nature \cite{das2023search}. To circumvent these problems, the modern search for MZMs focuses primarily on proximitized systems \cite{fu2008,Sau2010Generic, Lutchyn2010Majorana, Oreg2010Helical, Alicea2010Majorana}, using which various groups claimed to have observed MZMs due to the presence of zero-bias conductance peaks \cite{mourik2012signatures, das2012zero, deng2012anomalous, Finck2013Anomalous, Churchill2013Superconductor, nadj2014observation,chen2017experimental, vaitiekenas2021zero, Wang2022Plateau}. However, it is now clear that disorder-induced Andreev bound states can masquerade as MZMs in conductance experiments  \cite{Liu2012Zero,Roy2013Topologically,Chen2019Ubiquitous,prada2020,Pan2020Physical,Pan2021Quantized,Das2021Disorder,Vetal22}. Therefore, quenching disorder is an important goal in the pursuit of MZMs. 

As an exceptionally low-disorder platform \cite{Peres2010Colloquium,Sarma2011Electronic}, graphene is a  promising material for the realization of MZMs.\ Up to now, graphene-based proposals have involved the proximity effect due to the lack of  intrinsic superconductivity \cite{sanjose2015,penaranda2023majorana,xie2023gate}.\ Several  recent groundbreaking experiments shifted this paradigm by showing that graphene multilayers are robust, highly-tunable superconductors \cite{cao2018unconventional, yankowitz2019, lu2019superconductors,park2021tunable, hao2021electric,zhou2021superconductivity,park2022robust,zhou2022isospin,zhang2022promotion,zhang2023spin,holleis23Ising}, with mounting evidence suggesting that these states involve an exotic, non $s$-wave pairing \cite{oh2021evidence,kim2022evidence,lin2022zero,cao2021pauli, zhou2021superconductivity,park2022robust,zhou2022isospin}.\ In particular, spin-triplet, valley-odd $f$-wave pairing is emerging as a leading candidate for the superconducting symmetry \cite{oh2021evidence,kim2022evidence,sainzcruz2023junctions,CCFG22}. 
Inspired by these recent developments, we study models of \textit{intrinsic} $f$-wave superconductivity in chirally-stacked multilayer graphene. We show that in odd-layer configurations, these systems are topological mirror superconductors characterized by a nontrivial mirror winding number. Therefore, nanoribbons with mirror-symmetric edges \textit{must} host one Majorana mirror pair per boundary. Finite flakes also support zero modes with mirror character and degeneracy determined by the precise termination. Boundary states and their consequences for spectroscopy for the monolayer case were also considered in Refs. \cite{kokkeler23spectroscopic, Ghadimi2023Boundary}. Furthermore, we calculate the spectrum of finite flakes that host vortices. We find that a pair of zero modes is confined to each vortex core. This observation holds significant experimental implications since vortices can be readily created and probed by existing techniques \cite{SGRV14,Betal17,Ketal19,Cetal21,Zetal21,Letal23, Getal16,PNB19,Netal21}. Finally, we extend these results to twisted systems, wherein robust zero modes are found in twisted trilayer graphene, but not in twisted bilayer graphene. Importantly, a pair of zero modes is also trapped at vortices in twisted trilayer graphene. Our work suggests that multilayer graphene is an experimentally-feasible, ultra-clean Majorana platform.

\begin{figure}
    \centering
    \includegraphics[width=3.3in]{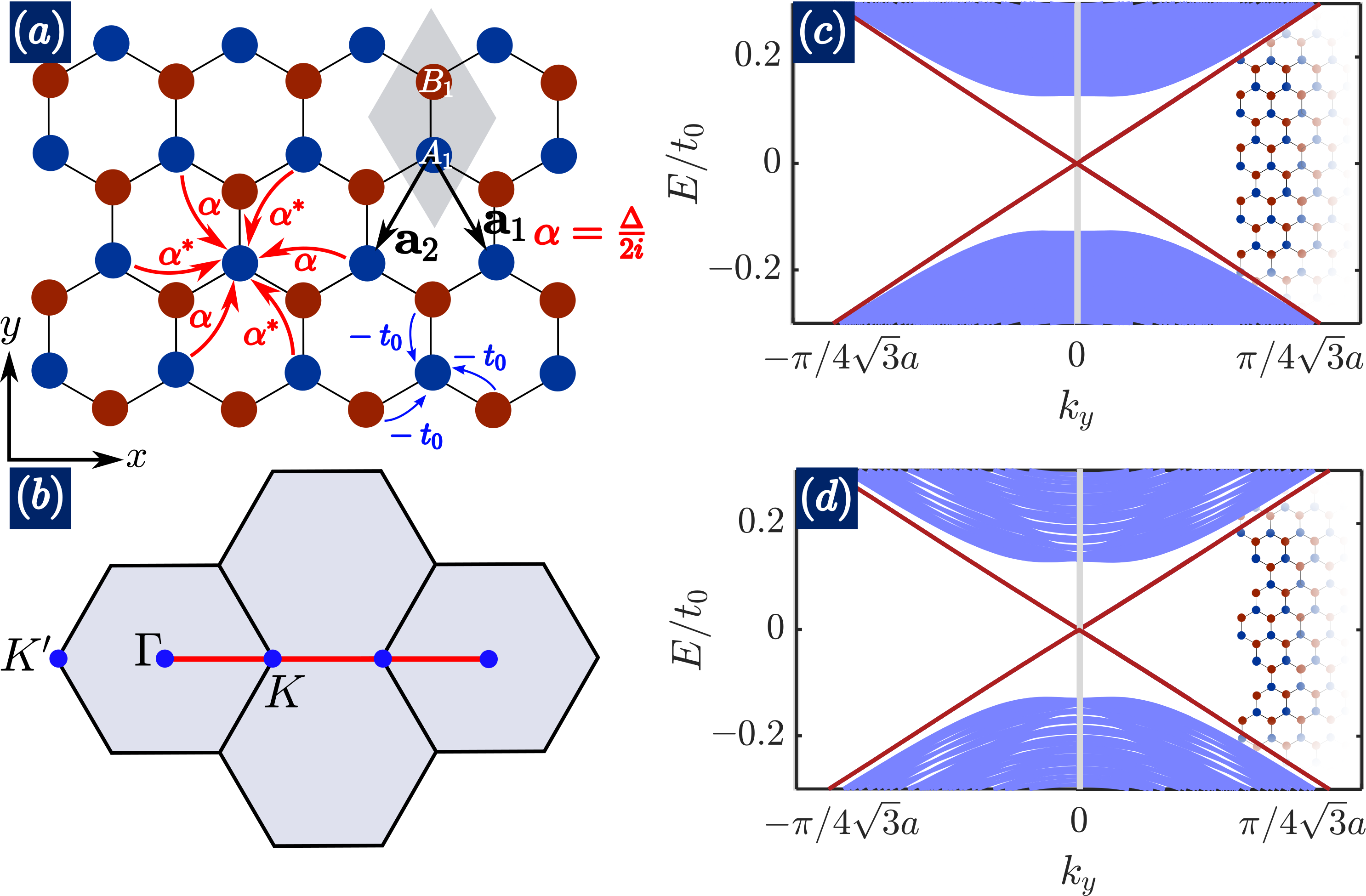}
    \caption{\textbf{Majorana  zero modes in a monolayer model.} (a) Tight-binding model of superconducting monolayer graphene. The hoppings between electrons and electrons (and between holes and holes) are indicated in blue, while the ``hoppings" between electrons and holes are indicated in red. (b) The Brillouin zone showing the mirror-symmetric line in red along which the winding number is calculated. (c)-(d) Band structures of mirror-symmetric nanoribbons showing protected edge states that cross zero energy at $k_y = 0.$ The edge structures are shown in the insets. Here, $\Delta = \mu = 0.05t_0,$ and the ribbons are 99a wide. }
    \label{fig:fig1}
\end{figure}

We begin with a simple model of \textit{spinless} superconducting monolayer graphene wherein the superconducting order parameter in real space is described by hoppings between the electrons and holes of the same sublattice with phase winding as indicated in Fig. \ref{fig:fig1}(a). Importantly, the order parameter changes signs between the $K$ and $K'$ valleys, realizing an exotic $f$-wave superconductor. We will show that at mirror-symmetric boundaries of nanoribbons, this model supports topologically-protected pairs of MZMs. The tight-binding Hamiltonian is described in Fig. \ref{fig:fig1}(a).  The Bogoliubov-de Gennes (BdG) Hamiltonian in momentum space is given by $\hat{\mathcal{H}} = \frac{1}{2} \sum_{\mathbf{k}} \hat{\Psi}^\dagger(\mathbf{k}) \mathcal{H}^\text{BdG}(\mathbf{k}) \hat{\Psi}(\mathbf{k}),$ where
\begin{equation}
\begin{split}
    \mathcal{H}^\text{BdG}(\mathbf{k}) & =  \tau_z \left[ \sigma_0 \mu + \sigma_xh_r(\mathbf{k})  -\sigma_yh_i(\mathbf{k}) \right]+  \tau_x \sigma_0 g(\mathbf{k}),\\
    h(\mathbf{k}) &= h_r(\mathbf{k})+ ih_i(\mathbf{k}) =  -t_0 \left(1+e^{i\kappa_1}+ e^{i \kappa_2}\right),  \\
    g(\mathbf{k}) &= \Delta \left[\sin \left( \kappa_1-\kappa_2 \right)+\sin \left( -\kappa_1 \right) +\sin \left( \kappa_2 \right) \right],
\end{split}
\end{equation}
$\kappa_i = \mathbf{k} \cdot \mathbf{a}_i,$ $\mathbf{a}_i$ are the lattice vectors, the $\tau$ and $\sigma$ Pauli matrices act on the Nambu particle-hole and sublattice $\lbrace A, B \rbrace$ spaces respectively with the subscript $0$ denoting the identity operator, and $\hat{\Psi}^\dagger(\mathbf{k}) = \left(\hat{c}^\dagger_{A,\mathbf{k}},\hat{c}^\dagger_{B,\mathbf{k}}, \hat{c}_{A,-\mathbf{k}}, \hat{c}_{B,-\mathbf{k}} \right).$ This Hamiltonian depends only on three real parameters: $t_0 > 0$ the hopping between nearest neighbors, $\Delta$ the superconducting parameter, and $\mu$ the chemical potential. Assuming that $\Delta \neq 0,$ the bulk band structure is gapped for $|\mu| < t_0$ and $|\mu| > 3 t_0$ \footnote{The latter condition is trivially true because it simply corresponds to fully-filled or fully-empty fermionic bands.}. We focus only on the lightly-doped regime.  Our system respects time reversal $\mathcal{T},$ particle hole $\mathcal{P},$ and mirror $\mathcal{M}_y$ symmetries:
\begin{equation}
    \begin{split}
        \mathcal{T} &= \tau_z \sigma_0 \mathcal{K}: \quad \mathcal{T}\mathcal{H}^\text{BdG}(\mathbf{k})\mathcal{T}^{-1} = \mathcal{H}^\text{BdG}(-\mathbf{k}), \\
        \mathcal{P} &= \tau_x \sigma_0 \mathcal{K}: \quad \mathcal{P}\mathcal{H}^\text{BdG}(\mathbf{k})\mathcal{P}^{-1} = -\mathcal{H}^\text{BdG}(-\mathbf{k}), \\
        \mathcal{M}_y &= \tau_0 \sigma_x: \quad \mathcal{M}_y\mathcal{H}^\text{BdG}(\mathbf{k})\mathcal{M}_y^{-1} = \mathcal{H}^\text{BdG}(\mathcal{M}_y \mathbf{k}),\\
    \end{split}
\end{equation}
where $\mathcal{K}$ is the complex conjugation operator. Combining $\mathcal{P}$ and $\mathcal{T}$  leads to a chiral symmetry $\mathcal{C} $
\begin{equation}
    \mathcal{C} = i\tau_y \sigma_0: \quad \mathcal{C}\mathcal{H}^\text{BdG}(\mathbf{k})\mathcal{C}^{-1} = -\mathcal{H}^\text{BdG}(\mathbf{k}),
\end{equation}
which requires that every state at $E(\mathbf{k})$ must have a partner at $-E(\mathbf{k}).$ Since both $\mathcal{T}^2 = \mathcal{P}^2 = +1,$ our system belongs to the topologically-trivial two-dimensional $BDI$ class \cite{Chiu2016Classification}. Therefore, a generic termination does not guarantee the existence of boundary states.

With mirror symmetry, we can enrich the topological classification along mirror-symmetric lines in the Brillouin zone  \cite{ZKM13a, ZKM13b, UYTS13, Shiozaki2014Topology}. In our case, there is only one independent mirror-symmetric line in the Brillouin zone along $k_y = 0$ as shown in Fig. \ref{fig:fig1}(b) \footnote{It may appear that there are two mirror-symmetric lines: one along $k_y = 0$ and one along $k_y = 2\pi/\sqrt{3}a,$ but these two are actually connected to each other by a reciprocal lattice vector. So they are \textit{not} independent.}. On this line $\left[ \mathcal{M}_y, \mathcal{H}^\text{BdG}(k_x,k_y=0) \right] = 0,$ so we can block-diagonalize the Hamiltonian into a mirror-odd and mirror-even sector.  Within each mirror sector, we can further put the Hamiltonian into chiral off-diagonal form since $\left[ \mathcal{M}_y,\mathcal{C} \right] = 0$ \cite{Chiu2016Classification,neupert2018topological, lin2021real},
\begin{equation}
    \mathcal{H}^\text{BdG} = \begin{pmatrix}
    \mathcal{H}_-& 0 \\
    0 & \mathcal{H}_+
    \end{pmatrix}, \quad \text{where} \quad \mathcal{H}_\pm = \begin{pmatrix}
    0 & \mathcal{D}_\pm  \\
    \mathcal{D}_\pm^\dagger  & 0
    \end{pmatrix}.
\end{equation}
The mirror-projected Hamiltonians $\mathcal{H}_\pm$ are characterized by winding numbers $\nu_\pm = \frac{1}{2\pi i} \oint dk_x \text{Tr} \left[\tilde{\mathcal{D}}_\pm^\dagger(k_x)\partial_{k_x} \tilde{\mathcal{D}}_\pm(k_x) \right],$ where $\tilde{\mathcal{D}}_\pm(k_x)$ is the obtained from $\mathcal{D}_\pm(k_x)$ via singular value decomposition \cite{neupert2018topological, lin2021real}. For our model, we have $\mathcal{D}_\pm (k_x) = \mu \pm h(k_x) - i g(k_x)$ and    $\nu_\pm = \mp \text{sign} \left( \Delta /t_0\right).$ This unity winding number predicts that when a mirror-symmetric edge is cut parallel to the $y$-direction, there exists one topologically-protected boundary mode per mirror sector at $k_y = 0.$ Importantly, this mode must reside exactly at zero energy due to chiral symmetry. More generally, the odd parity of $\nu_\pm$ guarantees the existence of at least one exact zero mode at $k_y = 0$. We illustrate these zero modes for two mirror-symmetric boundaries in Fig. \ref{fig:fig1}(c)-(d). In Fig. \ref{fig:fig1}(c), the termination is pristine armchair, while in Fig. \ref{fig:fig1}(d), the termination is jagged armchair that includes both armchair and zigzag characters. In both cases, since $\mathcal{M}_y$ is preserved, we find two zero modes, one from each mirror sector.

The Majorana zero modes can be  obtained analytically from a continuum theory for armchair edges. Assuming that $|\Delta|,|\mu| \ll t_0,$ the relevant physics is described by a Dirac theory in the original sublattice basis as
\begin{equation}
\label{eq: continuum Hamiltonian}
    \mathcal{H}^\text{BdG}(\mathbf{r}) = \tau_z \left[ \mu\sigma_0 -i \hbar v_F \nabla_\mathbf{r} \cdot \left( \xi \sigma_x, \sigma_y \right) \right] -  \xi \tilde{\Delta} \tau_x \sigma_0,
\end{equation}
where $\hbar v_F = \sqrt{3}t_0a/2,$ $\tilde{\Delta} = 3\sqrt{3} \Delta/2 ,$ and $\xi = \pm$ denotes valley. Putting the edge at $x = 0$ and extending into $x \rightarrow - \infty,$ the (unnormalized but normalizable) zero-energy solutions at $k_y = 0$ that satisfy the armchair boundary conditions are
\begin{equation}
\label{eq: analytic zero mode}
    \psi_{\xi,m_y} = e^{ x/\ell -iq_x x} \xi\begin{pmatrix} 1, & m_y, & -im_y\text{sign}\Delta, & - i \text{sign}\Delta \end{pmatrix}^T,
\end{equation}
where the decay length is  $\ell = \hbar v_F/|\tilde{\Delta}|,$ the wavelength is $q_x = m_y\xi \mu/\hbar v_F,$ and $m_y = \pm 1$ is the mirror eigenvalue of the mode. The valley-antisymmetric nature of the superconducting gap is crucial to the existence of the zero modes in Eq. \eqref{eq: analytic zero mode} because had the gap been endowed with the same sign in both valleys, these modes would not have been normalizeable. Thus, the presence of armchair-confined Majorana states can serve as a diagnostic of valley-odd, spin-triplet superconductivity in graphene \cite{kokkeler23spectroscopic}. Away from $k_y = 0,$ the mirror-odd and even sectors hybridize, lifting the energy away from zero. To find the dispersion, we can rewrite Eq. \eqref{eq: continuum Hamiltonian} at finite small $k_y$ in the basis of Eq. \eqref{eq: analytic zero mode}
\begin{equation}
    \mathcal{H}^\text{BdG}(k_y) = \hbar \tilde{v}_F \begin{pmatrix}
        0 & i k_y \\
        -i k_y & 0 
    \end{pmatrix}.
\end{equation}
Thus, the mid-gap boundary-projected dispersion is $\mathcal{E} = \pm \hbar \tilde{v}_F k_y,$ where $ \tilde{v}_F = v_F[1+(\mu/\tilde{\Delta})^2]^{-1}$, and the associated eigenfunctions are $\psi_{\xi, \mathcal{E} = \pm} = e^{i k_y y} \left[ \mp i \psi_{\xi,m_y =-1} +  \psi_{\xi,m_y =+1 } \right].$

For potential applications in topological quantum computing, it is desirable to isolate Majorana zero modes spatially. For our model, the Majorana zero modes always come in pairs as demanded by $\mathcal{M}_y$ along translationally-invariant edges. One way to isolate a Majorana zero mode is to break $\mathcal{M}_y$ symmetry by adding a spatially-dependent $\sigma_z$ mass of the form $V(x,y) = m(y) \tau_z \sigma_z$ where $\lim_{y\rightarrow\pm \infty} m(y) = \pm m$ and $m$ is a constant. The mass changes sign at $y=0.$ Then, there are two possible zero-energy solutions $\phi_{\xi,m_y} = e^{m_y\int^y m(y') dy'/\hbar v_F}\psi_{\xi,m_y}$ \footnote{We continue to label the $\phi_{\xi,m_y}$ states with mirror eigenvalue $m_y$ only as a matter of convenience. The reader should \textit{not} take this as implying that mirror symmetry is preserved. It is indeed broken explicitly by the potential $V(x,y)$}. However, only one of them is normalizable depending on whether $m$ is positive or negative. Thus, the effect of the mirror-breaking mass is to gap out the two boundary modes and create a single bound state localized  where the mass changes sign.

\begin{figure}
    \centering
    \includegraphics[width=3.3in]{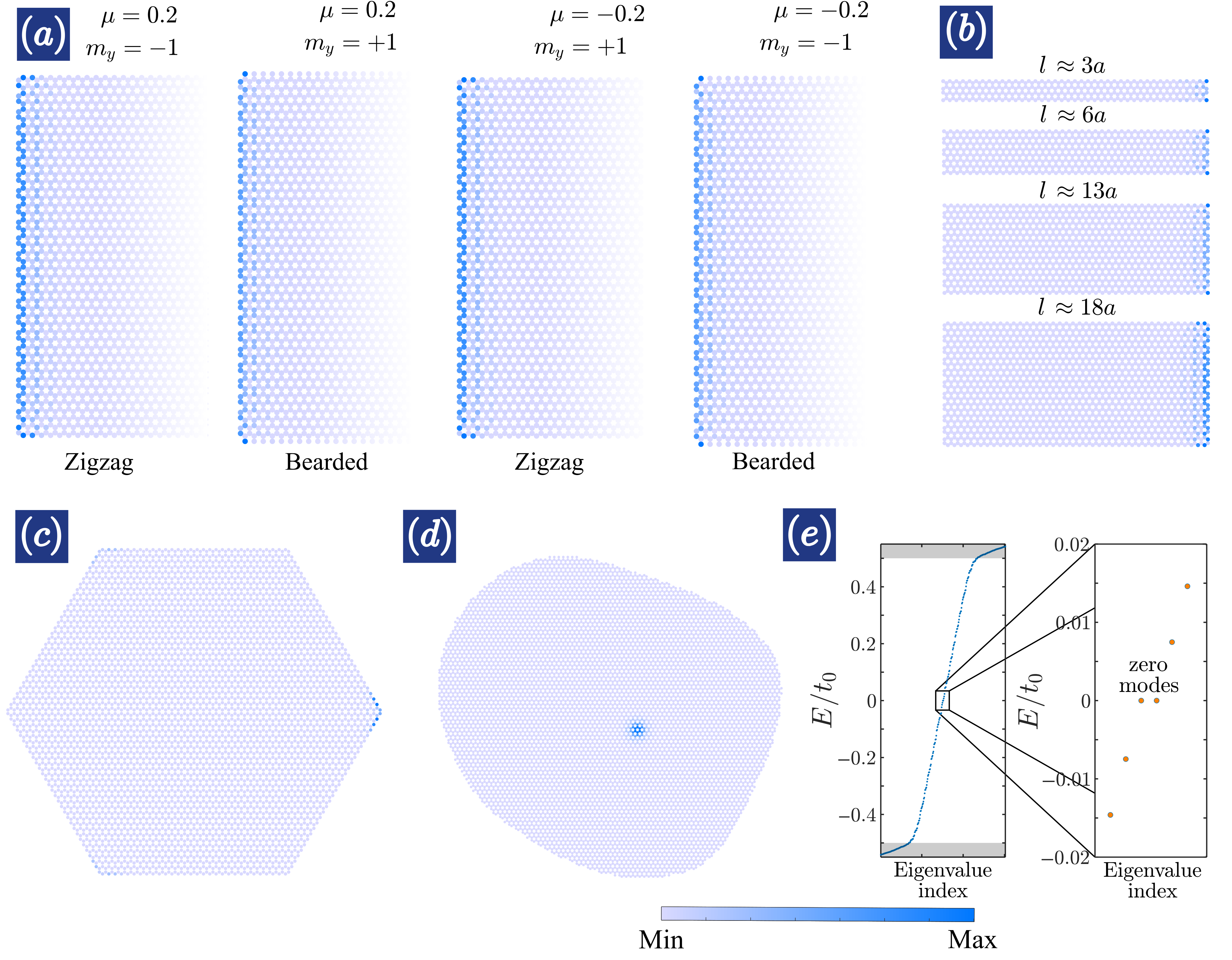}
    \caption{\textbf{Isolating Majorana zero modes on finite flakes.} (a) Majorana zero mode localized on the left armchair edge of a rectangular flake with different values of the chemical potential and different terminations of the horizontal top and bottom edges (either zigzag or bearded). The mirror eigenvalue of the shown state, $m_y,$ is indicated. (b) Dependence of the zero mode on vertical length of the flake. (c) Majorana corner state in a hexagonal flake. There are six such states in total. In these simulations, $\Delta = 0.1t_0$ and $|\mu| = 0.2t_0.$ (d) One of two vortex states confined to the  vortex core near zero energy $E \approx 10^{-17} t_0$, with $\Delta = 0.2t_0,$  $\mu = 0.1t_0,$ and $d \approx 0.$ The irregular shape has $9,609$ atoms, and is chosen to illustrate that the vortex states do not depend on boundary conditions. The energies of this flake are shown in (e), with the zero modes emphasized. The vortex is centered at a hexagon's center.   }
    \label{fig:2}
\end{figure}

Another way to isolate single  zero modes is by careful design of edges and corners \cite{Ghadimi2023Boundary}. For instance, in a rectangular flake that features armchair edges on the left and the right and either zigzag or bearded edges on both the top and the bottom, there are two zero modes that are localized on the left and right armchair edges. These modes are odd (even) under mirror if $\mu > 0$ and the horizontal edges are zigzag (bearded) or if $\mu < 0$ and the horizontal edges are bearded (zigzag), as shown in Fig. \ref{fig:2}(a). Therefore, the imposition of zigzag and bearded edges mimics a large $\sigma_z$ mass above the sample that goes to zero inside the sample and then changes sign below the sample. While the width separating the two armchair edges needs to be large to prevent mixing of the zero modes, the length of the flake along the $y$ direction does not qualitatively affect these modes, as shown in Fig. \ref{fig:2}(b). Furthermore, in a hexagonal flake, we find six localized corner states at zero energy, one of which is shown in Fig. \ref{fig:2}(c) \cite{Ghadimi2023Boundary}\footnote{The hexagonal shape can be thought of as an example of the rule that most clean, locally straight graphene edges are described, in the continuum limit, by the zigzag boundary conditions. This arises from the fact that the projection of the $K$ and $K'$ points on the edge only coincide (realizing the armchair boundary conditions) for very specific angles between the edges and the lattice axes \cite{AB08}. Edge regions connecting $A-$ and $B-$ like boundaries must include armchair like segments, where zero modes will be localized. Examples of zero modes in a distorted hexagon are shown in Ref. \cite{SM}.}. Spin domain walls can also host pairs of MZMs, see Ref. \cite{SM}. It is worth emphasizing that the chosen simulation parameters are unrealistically large to facilitate fast numerical convergence, but should not qualitatively affect the topologically-derived conclusions.

Crucially, vortices can also support zero-energy modes. Since vortices break translational symmetry, we calculate these vortex states numerically in real space on finite flakes using the tight-binding framework, where the gap function is modified to
$\Delta(\mathbf{r}_i,\mathbf{r}_j) = \pm e^{i \phi_\mathbf{r}} \Delta \tanh(|\mathbf{r}|/d)/2i,$  $\mathbf{r} = (\mathbf{r}_i+\mathbf{r}_j)/2$ is the center-of-mass position, $\phi_\mathbf{r}$ is the angular coordinate of $\mathbf{r},$ and $d$ is the coherence length. When the vortex is centered along a $\mathcal{M}_y$-invariant line that connects a hexagon's center and a mid-bond, we find two vortex-confined modes at numerically-exact zero energies, as shown in Fig. \ref{fig:2}(d). This is strong evidence that these vortex-confined zero modes are protected by mirror symmetry. When the vortex is centered elsewhere, such as on a carbon site, then these low-energy modes are no longer at zero energy.  However, when $d \gtrsim a,$ where $a$ is the lattice constant, the precise origin of the vortex becomes less important, and we find near-zero modes for various geometries numerically. The coherence length is the ratio of the Fermi velocity and the superconducting gap, which for practically-relevant small gaps, is many times the lattice constant. We leave the topological stability of these modes, including in the presence of disorder, to future analyses.

We now generalize the previous results to rhombohedral $N$-layer graphene stacks, which is motivated by the recently-discovered superconductivity in Bernal bilayer graphene and rhombohedral trilayer graphene \cite{zhou2021superconductivity,zhou2022isospin,zhang2023spin,holleis23Ising}. The Hamiltonian is now modified to 
\begin{equation}
    \mathcal{H}^\text{BdG}(\mathbf{k}) = \tau_z  \mathcal{H}_N(\mathbf{k}) + \tau_x  g(\mathbf{k}),
\end{equation}
where 
\begin{equation}
    \begin{split}
        \mathcal{H}_N(\mathbf{k}) &= \mathbb{I}_{N} \mathcal{H}_\text{intra}(\mathbf{k})  + \mathbb{U}_{N}  \mathcal{H}_\text{inter}(\mathbf{k}) + \mathbb{L}_{N}  \mathcal{H}^\dagger_\text{inter}(\mathbf{k}),  \\
        \mathcal{H}_\text{intra}(\mathbf{k}) &= \begin{pmatrix}
        \mu & h(\mathbf{k}) \\
        h^*(\mathbf{k}) & \mu
        \end{pmatrix}, \quad 
        \mathcal{H}_\text{inter}(\mathbf{k}) = \begin{pmatrix}
        0 & 0 \\
        \gamma_1 & 0
        \end{pmatrix},
    \end{split}
\end{equation}
$\mathbb{I}_{N}$ is the $N\times N$ identity matrix, $\mathbb{U}_{N}$  ($\mathbb{L}_{N} $) is the $N\times N$ matrix with ones along the diagonal above (below) the principal diagonal, acting on layer space, and $\gamma_1$ is the interlayer hopping.  For multiple layers, mirror symmetry is actually $\mathcal{C}_{2x}$ symmetry that simultaneously exchanges layers and sublattices represented by $\mathcal{M}_y = \tau_0\bar{\mathbb{I}}_N \sigma_x,$ where $\bar{\mathbb{I}}_{N}$ is the $N\times N$ \textit{anti}-diagonal matrix of ones. For $N$ even, all layers are exchanged under mirror, while for $N$ odd, there is one central layer which is mapped onto itself under mirror. Due to this mirror symmetry, we can classify the bands by the same topological invariant as before. For $N$ even, the system is topologically trivial with $\nu_\pm = 0.$ Consequently, there are no Majorana zero modes pinned to $k_y = 0$ on armchair nanoribbons. Interestingly, we find that for the current models, there are still zero modes displaced away from $k_y = 0.$ However, it is important to emphasize that these $k_y \neq 0$ modes are \textit{not} topologically protected, and thus they can be removed by perturbations which are invariant under $\mathcal{M}_y$ and do not close the bulk gaps. One such perturbation is a layer-antisymmetric $\sigma_z$ potential energy.

On the contrary, for $N$ odd, the systems are topological, protected by $\mathcal{M}_y$, with $\nu_\pm = \pm 1.$ Therefore, there are two topological zero modes per armchair edge pinned at $k_y = 0,$ in addition to many removable accidental zero modes at nonzero momenta. In addition to these zero modes, we also find numerous mid-gap states for $N>1.$ Like the non-topological zero modes, these mid-gap states are also removable; however, they are generically present at armchair edges. The topological zero modes can  be removed by breaking $\mathcal{M}_y$. In the multilayer case, one can break $\mathcal{M}_y$ by invalidating layer equivalence with a perpendicular electric field, which is much more experimentally accessible than a staggered potential. Regarding vortices, for the trilayer case, we have also found numerically a number of low-energy states confined to a vortex core. In fact, we find a pair of numerically-exact zero modes when the vortex center is located along a $\mathcal{M}_y$-symmetric line. Like before, if $d \gtrsim a,$ then the vortex center is not as important, so we find these near-zero modes to be robust. We expect all the odd-layer configurations to feature vortex-confined zero modes, but we have not numerically calculated these due to increasing computational cost.


\begin{figure}[t!]
    \centering
    \includegraphics[scale=0.37]{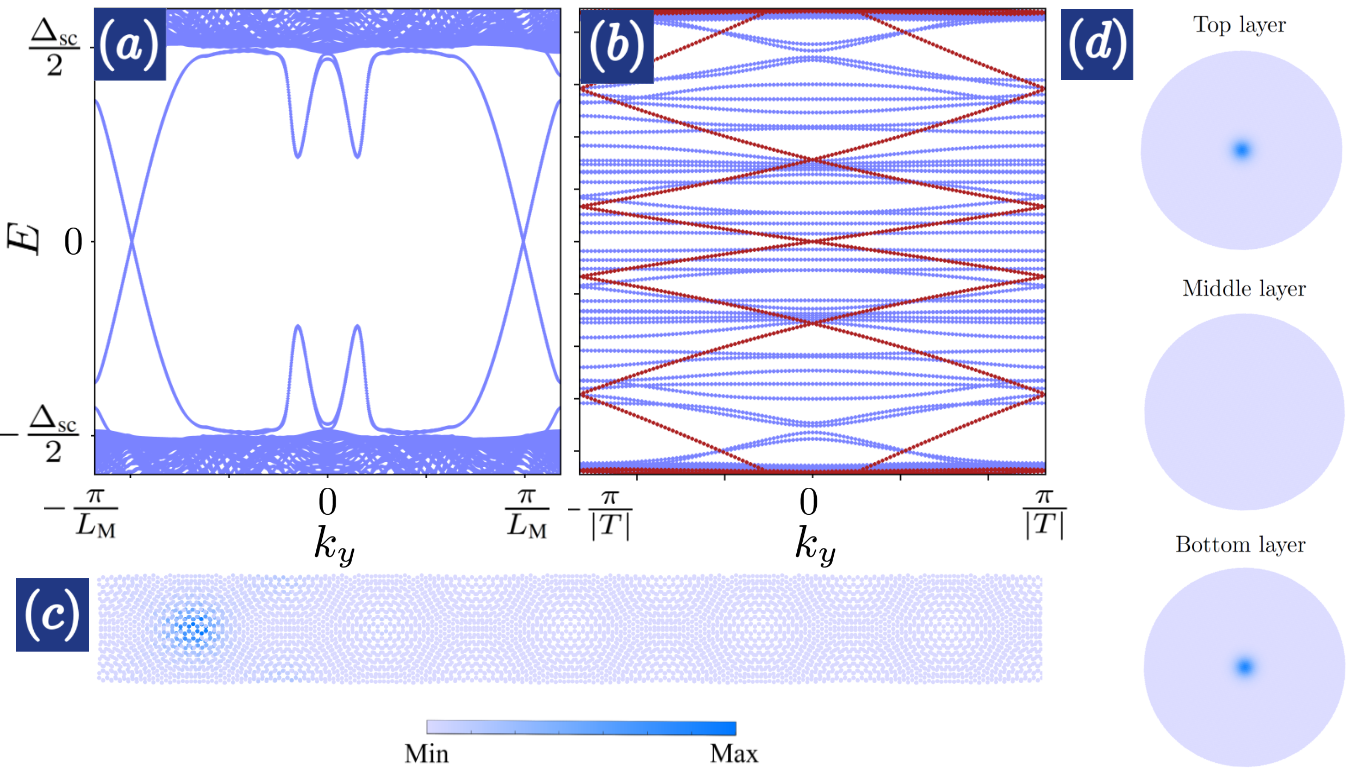}
    \caption{\textbf{Twisted multilayer graphene.} (a) Band structure of a twisted bilayer graphene nanoribbon, in the hole superconducting dome (filling $n=-2.4$), with a twist angle of $\theta=1.06^{\circ}$. The width of the ribbon is $W\approx35$ moiré periods ($L_M$) and the superconducting gap is $\Delta_\text{sc}=1$ meV. (b) Band structure of a twisted trilayer graphene nanoribbon with armchair edges in the top and bottom layers. The states in red come from the effective monolayer sector and thus have all the charge in top and bottom layers. Here $\theta=5^{\circ}$, $\mu=0$, $\Delta_\text{sc}=200$ meV, \cite{TTGparameters}, and $|T|>L_M$ due to a commensurability condition \cite{SM}. (c) Charge map of a TBG zero mode, showing localization at the last full moiré AA region near an edge. (d) Zero mode confined to a vortex core in twisted trilayer graphene, with $E \approx 10^{-10}t_0$. Here $\theta=1.5^{\circ}$, $n=-2.4$, $\Delta_\text{sc}=0.05t_0$, and the flake contains $323,742$ atoms.}
    \label{fig:TBGMZM}
\end{figure}

The above results can be generalized to $ABA$ stacks as well. When the number of layers is odd, such a stack can be decomposed into a direct sum of bilayer sectors and one monolayer sector due to the presence of $\mathcal{M}_z$ symmetry \cite{khalaf2019magic}. The monolayer sector behaves \textit{exactly} like the monolayer toy model albeit it is written in a basis of a coherent superposition of layers. As such, in these systems, there are also protected Majorana zero modes in the monolayer sector at armchair boundaries and robust vortex-confined low-energy modes. Although superconductivity has not yet been experimentally observed in $ABA$ multilayer graphene, these observations suggest that twisted multilayer graphene will contain the same exotic physics owning to the decomposition into a monolayer sector and bilayer sectors.

Inspired by the aforementioned findings, we  now look for zero modes in twisted multilayer graphene, using a scaled tight-binding model~\cite{gonzalez2017,lin2018minimum, vahedi2021magnetism,sainzcruz21high,sainzcruz2023junctions,guinea18electrostatic, bistritzer2011moire,  moon2014optical, SM}. Results for twisted bilayer and trilayer graphene appear in Fig. \ref{fig:TBGMZM}. In panel (a), the band structure of a TBG nanoribbon is shown for  angle $\theta = 1.06^\circ,$ which is close to the magic angle. The bands are qualitatively similar to those of Bernal bilayer graphene. There are 8 zero modes at $k_y\neq0$, which are \textit{not} topologically protected. These modes and the rest of sub-gap Andreev states appear near the edges of the system, at the last complete moiré AA regions, as shown in Fig. \ref{fig:TBGMZM}(c), and can be observed via Scanning Tunneling Microscopy \cite{oh2021evidence,kim2022evidence}.\ The bands depend sensitively on twist angle. Increasing the angle decreases the momentum of the near-zero modes.
When the angle is increased further, the zero modes disappear and only Andreev states near $k_y=0$ remain. Few sub-gap states survive at $\theta\gtrsim1.2^{\circ}$ \cite{SM}.

Just like in TBG, in twisted trilayer graphene (TTG) \cite{park2021tunable,hao2021electric}, generic $\mathcal{M}_y$-symmetric edges do not lead to topologically-protected zero modes at $k_y=0$. However, a new possibility arises due to the decomposition of TTG into TBG plus monolayer graphene, as found in Ref. \cite{khalaf2019magic}. In particular, since the effective monolayer comes from the odd combination of top and bottom layers, when these two layers have armchair edges, the system includes the 4 zero modes at $k_y=0$, like monolayer graphene, as shown in Fig. \ref{fig:TBGMZM}(b). The charge of these zero modes is indeed evenly distributed in top and bottom layers, with no charge in the middle layer.\ Since a decomposition including a monolayer is possible for any alternating-twist stack with an odd number of layers \cite{khalaf2019magic}, we expect that 4 zero modes at $k_y=0$ will exist in all such multilayers. Finally, Fig. \ref{fig:TBGMZM}(d) shows one of the two zero modes confined to a vortex core in twisted trilayer graphene. As in the ribbon geometry, these zero modes come from the effective monolayer sector. These promising results suggest that there is a whole family of robust superconductors in twisted odd-layer graphene that host topologically-protected Majorana zero modes.

Superconductivity with $f$-wave pairing is favored in two-dimensional materials with Fermi pockets at opposing corners of the Brillouin zone and strong repulsive electron-electron interactions. Defects which scatter electrons between the different pockets lead, in the superconducting phase, to Andreev resonances deep within the gap, and even allow for the existence of isolated Majorana states. In particular, the Majorana states are present when $\mathcal{M}_y$ symmetry is respected. This symmetry can be broken by applying a perpendicular displacement field. In rhombohedral stacks, superconductivity has not yet been seen without such a field; so our model might not immediately apply to these cases \cite{zhou2021superconductivity,zhou2022isospin,zhang2023spin,holleis23Ising}. However, superconductivity has been unambiguously established in twisted trilayer graphene even in the absence of a displacement field \cite{park2021tunable, cao2021large, hao2021electric}. Therefore, twisted trilayer graphene is the most promising candidate to detect these Majorana bound states, the observation of which is a simple way to identify this exotic $f$-wave superconductivity. In these materials, edges and corners are prototypical defects that trap Andreev states and Majorana zero modes. Although precision fabrication of these boundaries is technically challenging, there has been much experimental progress in achieving such a feat \cite{cai2010atomically, talirz2016surface,ruffieux2016surface, kitao2020scalable, houtsma2021atomically}. On the other hand, vortices are much more experimentally accessible because they can be created in the bulk by a magnetic field or by magnetic impurities. The electronic states at vortex cores have been extensively studied using local probes \cite{SGRV14,Betal17,Ketal19,Cetal21,Zetal21,Letal23}. These techniques can also be used to manipulate vortices \cite{Getal16,PNB19,Netal21}. Therefore, the vortex-confined zero modes in these graphene-based topological superconductors can be created, probed, and manipulated by readily-available technology, opening the way to new types of quantum devices.

We thank Fernando de Juan and Pierre A. Pantale\'{o}n for insightful discussions. V. T. P. and E. J. M. are supported by  the U.S. Department of Energy under grant DE-FG02-84ER45118. IMDEA Nanociencia acknowledges support support support from the \textquotedblleft Severo Ochoa\textquotedblright~Programme for Centres of Excellence in R\&D (CEX2020-001039-S/AEI/10.13039/501100011033).\  H. S-C and F. G acknowledge funding from the European Commission, within the Graphene Flagship, Core 3, Grant No. 881603; and from grants NMAT2D (Comunidad de Madrid, Spain), SprQuMat and (MAD2D-CM)-MRR MATERIALES AVANZADOS-IMDEA-NC. NOVMOMAT, project PID2022-142162NB-I00 funded by MICIU/AEI/10.13039/501100011033 and by FEDER, UE.

\clearpage

\setcounter{equation}{0}
\setcounter{figure}{0}
\renewcommand{\theequation}{S\arabic{equation}}
\renewcommand{\thefigure}{S\arabic{figure}}
\renewcommand{\thetable}{S\roman{table}}
\renewcommand{\bibnumfmt}[1]{[#1]}
\renewcommand{\citenumfont}[1]{#1}

\onecolumngrid

\begin{center}
\begin{Large}
Supplementary Material 
\end{Large}
\end{center}

\section{Monolayer graphene toy model}

\subsection{Tight-binding mean-field Hamiltonian}

We consider the following spinless Hamiltonian defined on a bipartite honeycomb lattice
\begin{equation}
\label{eq: Hamiltonian}
    \hat{\mathcal{H}} = -t_0 \sum_{\langle i j \rangle} \hat{c}_i^\dagger \hat{c}_j + \mu \sum_{i} \hat{c}_i^\dagger \hat{c}_i  +\frac{\Delta }{4i}\sum_{\langle \langle ij \rangle \rangle } (-1)^{3\theta_{ij}} \hat{c}_i^\dagger \hat{c}_j^\dagger-\frac{\Delta }{4i} \sum_{\langle \langle ij \rangle \rangle } (-1)^{3\theta_{ij}} \hat{c}_j \hat{c}_i,
\end{equation}
where $t_0>0$ is the nearest-neighbor hopping constant, $\mu$ is the chemical potential, $\Delta$ is the magnitude of the superconducting order parameter, and $\theta_{ij}$ is the angle between $\mathbf{r}_j-\mathbf{r}_i$ and the $x$-axis in units of $\pi$. We note parenthetically that the particle non-conserving terms look similar to the Haldane next-nearest-neighbor hopping terms, except that there is no sublattice dependence in our case. The fermionic operators satisfy $\lbrace \hat{c}_i, \hat{c}_j^\dagger \rbrace = \delta_{ij}.$ To obtain the momentum-space Hamiltonian, we define
\begin{equation}
    \hat{c}^\dagger_\mathbf{k} = \frac{1}{\sqrt{\mathcal{N}}} \sum_{\mathbf{r}_i} e^{i\mathbf{k} \cdot \mathbf{r}_i} \hat{c}_i^\dagger, \quad \hat{c}_i^\dagger = \frac{1}{\sqrt{\mathcal{N}}} \sum_{\mathbf{k}} e^{-i\mathbf{k} \cdot \mathbf{r}_i}\hat{c}^\dagger_\mathbf{k},
\end{equation}
to find
\begin{equation}
\begin{split}
    \hat{\mathcal{H}} = &-t_0 \sum_{\mathbf{k}} \left( h_\mathbf{k}\hat{c}^\dagger_{\mathbf{k}, A} \hat{c}_{\mathbf{k},B} + h_\mathbf{k}^*\hat{c}^\dagger_{\mathbf{k}, B} \hat{c}_{\mathbf{k},A} \right) + \mu \sum_{\mathbf{k}} \left( \hat{c}^\dagger_{\mathbf{k}, A} \hat{c}_{\mathbf{k},A} + \hat{c}^\dagger_{\mathbf{k}, B} \hat{c}_{\mathbf{k},B}\right) + \\
                &+ \frac{\Delta}{2}\sum_{\mathbf{k}} g_\mathbf{k} \left( \hat{c}^\dagger_{\mathbf{k}, A} \hat{c}^\dagger_{-\mathbf{k},A} +\hat{c}^\dagger_{\mathbf{k}, B} \hat{c}^\dagger_{-\mathbf{k},B} \right) + \frac{\Delta}{2}\sum_{\mathbf{k}} g_\mathbf{k} \left(  \hat{c}_{-\mathbf{k},A}\hat{c}_{\mathbf{k}, A} + \hat{c}_{-\mathbf{k},B}\hat{c}_{\mathbf{k}, B} \right),
\end{split}
\end{equation}
where $h_\mathbf{k} = 1+e^{i\mathbf{k} \cdot \mathbf{a}_1}+ e^{i \mathbf{k} \cdot \mathbf{a}_2}$ and $g_\mathbf{k} = \sin \left(\mathbf{k} \cdot \left(\mathbf{a}_1-\mathbf{a}_2 \right) \right)+ \sin \left( -\mathbf{k} \cdot \mathbf{a}_1 \right)+ \sin \left( \mathbf{k} \cdot \mathbf{a}_2 \right).$ Now, using the commutation relation $\lbrace \hat{c}_{\mathbf{k},\sigma}, \hat{c}_{\mathbf{k}',\sigma'}^\dagger \rbrace = \delta_{\mathbf{k},\mathbf{k}'} \delta_{\sigma, \sigma'},$ we can rewrite this Hamiltonian as 
\begin{equation}
\begin{split}
    \hat{\mathcal{H}} = &-\frac{t_0}{2} \sum_{\mathbf{k}} \left( h_\mathbf{k}\hat{c}^\dagger_{\mathbf{k}, A} \hat{c}_{\mathbf{k},B} + h^*_\mathbf{k}\hat{c}^\dagger_{\mathbf{k}, B} \hat{c}_{\mathbf{k},A} \right)+ \frac{\mu}{2} \sum_{\mathbf{k}} \left( \hat{c}^\dagger_{\mathbf{k}, A} \hat{c}_{\mathbf{k},A} + \hat{c}^\dagger_{\mathbf{k}, B} \hat{c}_{\mathbf{k},B}\right) + \\
    &+\frac{t_0}{2} \sum_{\mathbf{k}} \left( h_{-\mathbf{k}}  \hat{c}_{-\mathbf{k},B}\hat{c}^\dagger_{-\mathbf{k}, A} +  h^*_{-\mathbf{k}}\hat{c}_{-\mathbf{k},A}\hat{c}^\dagger_{-\mathbf{k}, B}\right) - \frac{\mu}{2} \sum_{\mathbf{k}} \left(  \hat{c}_{-\mathbf{k},A}\hat{c}^\dagger_{-\mathbf{k}, A} + \hat{c}_{-\mathbf{k},B} \hat{c}^\dagger_{-\mathbf{k}, B}\right) + \\
    &+ \frac{\Delta}{2}\sum_{\mathbf{k}} g_\mathbf{k} \left( \hat{c}^\dagger_{\mathbf{k}, A} \hat{c}^\dagger_{-\mathbf{k},A} +\hat{c}^\dagger_{\mathbf{k}, B} \hat{c}^\dagger_{-\mathbf{k},B} \right) + \frac{\Delta}{2}\sum_{\mathbf{k}} g_\mathbf{k} \left( \hat{c}_{-\mathbf{k},A} \hat{c}_{\mathbf{k}, A} +\hat{c}_{-\mathbf{k},B}\hat{c}_{\mathbf{k}, B}  \right) + \text{constant.} 
\end{split}
\end{equation}
We henceforth drop the inconsequential constant term. Putting this into a convenient matrix form, we obtain, using $h_\mathbf{k} = h^*_{-\mathbf{k}},$
\begin{equation}
    \begin{split}
        \hat{\mathcal{H}} &= \frac{1}{2} \sum_\mathbf{k} \hat{\Psi}^\dagger(\mathbf{k}) \mathcal{H}^\text{BdG}(\mathbf{k})\hat{\Psi}(\mathbf{k}),\\
        \mathcal{H}^\text{BdG}(\mathbf{k}) &= \begin{pmatrix}
        \mu & -t_0 h_\mathbf{k} & \Delta g_\mathbf{k} & 0 \\
        -t_0 h^*_\mathbf{k} & \mu & 0 & \Delta g_\mathbf{k} \\
        \Delta g_\mathbf{k} & 0 & -\mu & t_0 h_\mathbf{k} \\
        0 & \Delta g_\mathbf{k} & t_0 h^*_\mathbf{k} & - \mu
        \end{pmatrix},\\
        \hat{\Psi}^\dagger(\mathbf{k}) &= \begin{pmatrix} \hat{c}_{\mathbf{k},A}^\dagger & \hat{c}_{\mathbf{k},B}^\dagger & \hat{c}_{-\mathbf{k},A} & \hat{c}_{-\mathbf{k},B} \end{pmatrix}.
    \end{split}
\end{equation}

\begin{figure}
    \centering
    \includegraphics[width=4in]{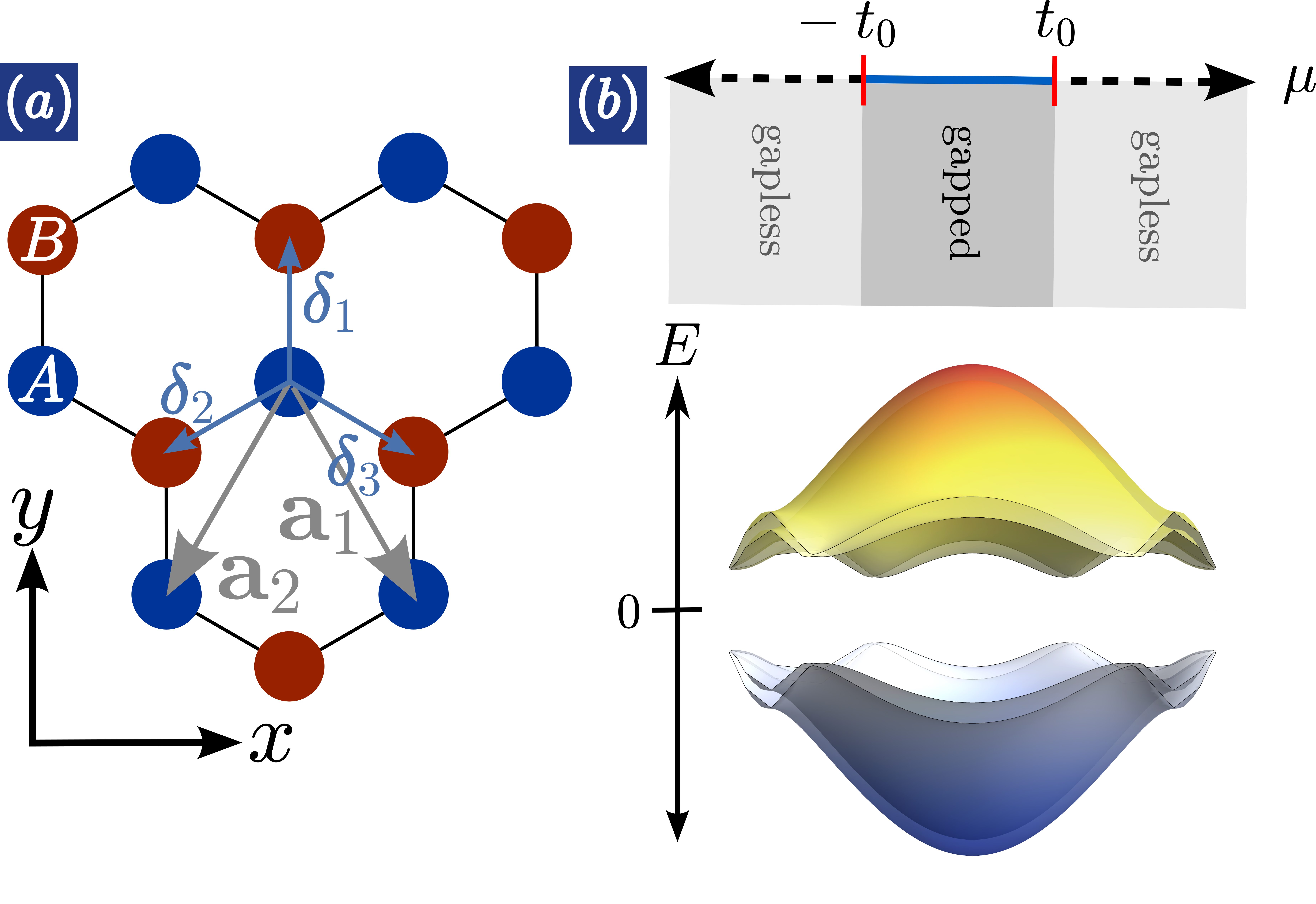}
    \caption{(a) Lattice structure, (b) phase diagram, and a representative band structure of the model Hamiltonian in Eq. \eqref{eq: Hamiltonian}. Here, we use $t_0 = 1,$ $\Delta = 0.2,$ and $\mu = 0.1$ for the gapped phase.}
    \label{fig:lattice}
\end{figure}

Next, we study the symmetries of Hamiltonian \eqref{eq: Hamiltonian}. Because of the superconducting order parameter, this Hamiltonian breaks $\mathcal{C}_{2z}$ rotation symmetry and $\mathcal{M}_x$ mirror symmetry. However, $\mathcal{C}_{3z}$ rotation symmetry and $\mathcal{M}_y$ mirror symmetry are preserved. These are implemented by unitary operators:
\begin{equation}
\begin{split}
 \mathcal{C}_{3z} : &\quad   \hat{\mathcal{C}}_{3z} \begin{pmatrix}\hat{c}^\dagger_A\left(\mathbf{r}_i\right) \\ \hat{c}_A\left(\mathbf{r}_i\right) \end{pmatrix}\hat{\mathcal{C}}_{3z}^{-1} = \begin{pmatrix} \hat{c}^\dagger_A\left(\mathcal{R}_{3z}\mathbf{r}_i\right)  \\ \hat{c}_A\left(\mathcal{R}_{3z}\mathbf{r}_i\right)\end{pmatrix},\\
    &\quad \hat{\mathcal{C}}_{3z}\begin{pmatrix}\hat{c}^\dagger_B\left(\mathbf{r}_i\right) \\ \hat{c}_B\left(\mathbf{r}_i\right) \end{pmatrix}\hat{\mathcal{C}}_{3z}^{-1} = \begin{pmatrix} \hat{c}^\dagger_B\left(\mathcal{R}_{3z}\mathbf{r}_i\right)  \\ \hat{c}_B\left(\mathcal{R}_{3z}\mathbf{r}_i\right)\end{pmatrix}        ,
\end{split}
\end{equation}
\begin{equation}
\begin{split}    
\mathcal{M}_y : &\quad   \hat{\mathcal{M}}_y \begin{pmatrix} \hat{c}_A^\dagger(\mathbf{r}_i) \\ \hat{c}_A(\mathbf{r}_i) \end{pmatrix} \hat{\mathcal{M}}_y^{-1} = \begin{pmatrix}\hat{c}_B^\dagger \left(\mathcal{M}_y\mathbf{r}_i\right) \\ \hat{c}_B \left(\mathcal{M}_y\mathbf{r}_i\right) \end{pmatrix},\\
    &\quad   \hat{\mathcal{M}}_y \begin{pmatrix}\hat{c}_B^\dagger(\mathbf{r}_i) \\ \hat{c}_B(\mathbf{r}_i) \end{pmatrix}\hat{\mathcal{M}}_y^{-1} = \begin{pmatrix}\hat{c}_A^\dagger \left(\mathcal{M}_y\mathbf{r}_i\right) \\\hat{c}_A \left(\mathcal{M}_y\mathbf{r}_i\right) \end{pmatrix},
\end{split}
\end{equation}
where we have adopted an alternative convention to $\hat{c}_i^\dagger$ to emphasize the spatial dependence of the creation operators so that $\hat{c}_\sigma^\dagger(\mathbf{r})$ creates a fermion on the $\sigma$ sublattice located at spatial point $\mathbf{r}.$ We note that while $\mathcal{C}_{3z}$ preserves sublattice, $\mathcal{M}_y$ exchanges sublattices. In momentum space, the mirror symmetry is represented by
\begin{equation}
\begin{split}
    \mathcal{M}_y : &\quad  \left(\tau_0 \otimes \sigma_x\right) \mathcal{H}^\text{BdG}(\mathbf{k}) \left(\tau_0 \otimes \sigma_x\right) = \mathcal{H}^\text{BdG}(\mathcal{M}_y\mathbf{k}).
\end{split}
\end{equation}
Here, the $\sigma$ Pauli matrices act on the sublattice degree of freedom and the $\tau$ Pauli matrices act on the Nambu space. One can easily verify that the Hamiltonian does indeed satisfy these constraints. We note that $\mathcal{M}_y^2 = +1.$ In addition to these spatial symmetries, there is also a local antiunitary particle-hole symmetry. It is most easily seen in first-quantized form in momentum space where it takes the form
\begin{equation}
    \mathcal{P}: \quad \left(\tau_x\otimes \sigma_0 \right)\left[\mathcal{H}^\text{BdG}(\mathbf{k})\right]^*\left(\tau_x\otimes \sigma_0 \right) = -\mathcal{H}^\text{BdG}(-\mathbf{k}).
\end{equation}
 Additionally, we observe that there is spinless time-reversal symmetry $\mathcal{T}$ represented by
\begin{equation}
    \mathcal{T}: \quad \left(\tau_z\otimes \sigma_0 \right)\left[\mathcal{H}^\text{BdG}(\mathbf{k})\right]^*\left(\tau_z\otimes \sigma_0 \right) = \mathcal{H}^\text{BdG}(-\mathbf{k}).
\end{equation}
because $-g_\mathbf{k} =  g_{-\mathbf{k}}$ but $g_\mathbf{k}^* = g_\mathbf{k},$ necessitating an extra minus sign in the off-diagonal elements in Nambu space. We note that $\mathcal{T}^2 = +1.$ Of course, the Hamiltonian also respects chiral symmetry $\mathcal{C} = \mathcal{T}\mathcal{P}$ since $\mathcal{P}$ and $\mathcal{T}$ are individually respected
\begin{equation}
    \mathcal{C}: \quad \left(i\tau_y\otimes \sigma_0 \right)\mathcal{H}^\text{BdG}(\mathbf{k})\left(-i\tau_y\otimes \sigma_0 \right) = -\mathcal{H}^\text{BdG}(\mathbf{k}).
\end{equation}

We now study the band structure of Eq. \eqref{eq: Hamiltonian}. The four energy bands of the BdG matrix are given by 
\begin{equation}
    E(\mathbf{k}) = \pm \sqrt{\Delta^2 g_\mathbf{k}^2 + \left( \mu \pm t_0 |h_\mathbf{k}|\right)^2}.
\end{equation}
Here, $g_\mathbf{k} = 2 \sin \left(k_xa/2\right) \left[\cos \left(k_xa/2\right)-\cos \left(\sqrt{3} k_ya/2\right)\right],$ which vanishes along the lines $k_x = 0, k_x = \pm \sqrt{3} k_y$ in the first Brillouin zone. Because of $C_{3z}$ rotation symmetry, we focus on the $k_x=0$ line. The other ones are related by symmetry. Along this line, $|h_{k_y}| = \sqrt{5+4\cos(\sqrt{3}k_ya/2)}.$ For $k_y \in \left[-2\pi /\sqrt{3}a, 2 \pi /\sqrt{3}a \right],$ the range of $\cos(\sqrt{3}k_ya/2)$ is $[-1,1].$ So $|h_{k_y}|$ varies continuously in $[1,3]$ as $k_y$ is sampled across the Brillouin zone. Therefore, always assuming $\Delta \neq 0,$ we have a gap closing at $E = 0$ if 
\begin{equation}
  t_0 \leq   \left|\mu\right| \leq 3t_0,
\end{equation}
and we have a gapped phase if
\begin{equation}
    \left|\mu\right| < t_0 \text{ or } \left|\mu\right| > 3t_0.
\end{equation}
Conventionally, the chemical potential will be small; so we will almost always be in a gapped phase. To access the gapless phase, we have to dope  at least to $\mu = \pm t_0.$ For $|\mu| > 3t_0,$ we technically have another gapped phase at $E = 0,$ but this is kind of a trivial  observation because it corresponds to fully empty or fully filled fermionic bands. So we will not consider these large values of $\mu$ further since they are irrelevant to our purpose. A representative band structure is shown in Fig.~\ref{fig:lattice}(b).


\subsection{Symmetry classification using the winding number}

Now that the spectral properties of the band structure in parameter space are well understood, we can classify the topology of the bulk bands below $E = 0$ in the case that the spectrum is gapped. First, since we have both $\mathcal{T}^2 = +1$ and $\mathcal{P}^2 = +1,$ the system  is in the two-dimensional symmetry class $BDI,$ which is topologically trivial according to the ten-fold classification \cite{CYR13}. However, we can enrich the classification by considering mirror symmetry. Due to the presence of $\mathcal{M}_y$ symmetry, we can label states by mirror eigenvalues $m_y=\pm 1$ along mirror-symmetric momentum lines \cite{ZKM13a, ZKM13b, UYTS13, CYR13}. Notice that the mirror eigenvalues are real because we are dealing with a spinless representation. Along each of these lines, the system is one-dimensional and admits a classification via the winding number within each mirror sector.

In our case, there is only one independent mirror-invariant line in the Brillouin zone along $k_y = 0.$ The line at $k_y = \frac{2\pi}{\sqrt{3}a}$ is connected to the line at $k_y=0$ via a reciprocal lattice vector so it cannot be considered an independent line. In fact, in order to traverse a closed loop in $\mathbf{k}$-space along a mirror-invariant line with $k_y = 0,$ we need to go from $k_x = 0$ to $k_x = 4\pi/a,$ which goes through both the line at $k_y=0$ and the line at $k_y=\frac{2\pi}{\sqrt{3}a},$ showing that they are indeed on the same line.  This is also clear from observation of the explicit form of the Hamiltonian 
\begin{equation}
\begin{split}
\mathcal{H}^\text{BdG}(k_x) &= \left(
\begin{array}{cccc}
 \mu  & h(k_x) & g(k_x) & 0 \\
 h(k_x) & \mu  & 0 & g(k_x) \\
 g(k_x) & 0 & -\mu  & -h(k_x) \\
 0 & g(k_x) & -h(k_x) & -\mu  \\
\end{array}
\right), \\
h(k_x) &= -t_0 \left(2 \cos \left(\frac{k_xa}{2}\right)+1\right), \\
g(k_x) &= \Delta  \left(\sin (k_xa)-2 \sin \left(\frac{k_xa}{2}\right)\right),
\end{split}
\end{equation}
which has period $4\pi/a.$ We have absorbed $t_0$ into $h$ and $\Delta$ into $g$ for brevity. Along the mirror-symmetric line, we have
\begin{equation}
\begin{split}
    \mathcal{M}_y\mathcal{H}^\text{BdG}(k_x,k_y = 0)\mathcal{M}_y^{-1} &= \mathcal{H}^\text{BdG}(k_x,k_y = 0).
\end{split}
\end{equation}
So we can partition the Hamiltonian into a mirror-even and mirror-odd block Hamiltonian
\begin{equation}
    \mathcal{H}^\text{BdG}(k_x) = \begin{pmatrix}
    \mathcal{H}_-(k_x) & 0 \\
    0 & \mathcal{H}_+(k_x)
    \end{pmatrix},
\end{equation}
where
\begin{equation}
\begin{split}
\mathcal{H}_-(k_x) &=\left(
\begin{array}{cc}
 h(k_x) - \mu & g(k_x) \\
 g(k_x) & -h(k_x)+\mu \\
\end{array}
\right),   \\
\mathcal{H}_+(k_x) &=\left(
\begin{array}{cc}
 -h(k_x) - \mu & g(k_x) \\
 g(k_x) & h(k_x) + \mu\\
\end{array}
\right).    
\end{split}
\end{equation}
The mirror basis consists of 
\begin{equation}
    \begin{split}
        m_y=\pm1: \quad \ket{1_\pm} = \frac{1}{\sqrt{2}}\begin{pmatrix}0 \\ 0\\ \pm1 \\ 1\end{pmatrix}, \quad \ket{2_\pm}  = \frac{1}{\sqrt{2}}\begin{pmatrix} \pm1\\ 1\\ 0\\ 0\end{pmatrix}. 
    \end{split}
\end{equation}
The subscript refers to the mirror sector (i.e. the mirror eigenvalue of that state), and the numbers $1,2$ just label the states within each mirror sector. Now because $[\mathcal{M}_y,\mathcal{C}] = 0,$ we use $\mathcal{C}$ also to simultaneously put the mirror-diagonal Hamiltonian into chiral form. In the mirror basis, the chiral operator is \begin{equation}
\begin{split}
    \mathcal{C} = \begin{pmatrix} \mathcal{C}_- & 0 \\
    0 & \mathcal{C}_+
    \end{pmatrix}, \quad 
    \mathcal{C}_- = \mathcal{C}_+ = \left(
\begin{array}{cc}
 0 & -1 \\
 1 & 0 \\
\end{array}
\right).    
\end{split}
\end{equation}
For instance, in the following basis within the mirror-odd sector
\begin{equation}
    m_y = -1: \quad \ket{-,+i} = \frac{1}{\sqrt{2}} \begin{pmatrix}i \\ 1\end{pmatrix}, \quad \ket{-,-i} = \frac{1}{\sqrt{2}} \begin{pmatrix} -i\\1\end{pmatrix},
\end{equation}
the mirror-odd Hamiltonian can be put into off-diagonal chiral form
\begin{equation}
\begin{split}
    \mathcal{H}_-(k_x) &= \begin{pmatrix}
    0 & \mathcal{D}_-(k_x) \\
    \mathcal{D}^\dagger_-(k_x) & 0
    \end{pmatrix}, \\
    \mathcal{D}_-(k_x) &= 2 i \Delta  \sin \left(\frac{k_xa}{2}\right)-i \Delta  \sin (k_xa)+2 t_0 \cos \left(\frac{k_xa}{2}\right)+\mu +t_0.
\end{split}
\end{equation}
In a similar manner, we can also rewrite the mirror-even Hamiltonian in the following basis
\begin{equation}
    m_y = +1: \quad \ket{+,+i} = \frac{1}{\sqrt{2}} \begin{pmatrix} i  \\ 1\end{pmatrix}, \quad \ket{+,-i} = \frac{1}{\sqrt{2}} \begin{pmatrix} -i \\ 1\end{pmatrix}, 
\end{equation}
so that it too is in chiral form
\begin{equation}
\begin{split}
    \mathcal{H}_+(k_x) &= \begin{pmatrix}
    0 & \mathcal{D}_+(k_x) \\
    \mathcal{D}^\dagger_+(k_x) & 0
    \end{pmatrix}, \\
    \mathcal{D}_+(k_x) &= 2 i \Delta  \sin \left(\frac{k_xa}{2}\right)-i \Delta  \sin (k_xa)-2 t_0 \cos \left(\frac{k_xa}{2}\right)+\mu -t_0.
\end{split}
\end{equation}
Next, we ``flatten" the Hamiltonian so that all the positive energy eigenvalues are $+1$ and all the negative energy eigenvalues are $-1.$ This is accomplished using the singular value decomposition \cite{lin2021real} 
\begin{equation}
    \mathcal{D}_\pm(k_x) = \mathcal{U}_\pm(k_x) \Lambda_\pm(k_x) \mathcal{V}^\dagger_\pm(k_x).
\end{equation}
The topological invariant is the winding number
\begin{equation}
    \begin{split}
       \nu_\pm(k_y=0) &= \frac{1}{2\pi i} \int_0^{4\pi/a} dk_x \text{Tr} \left(\tilde{\mathcal{D}}^\dagger_\pm(k_x) \partial_{k_x} \tilde{\mathcal{D}}_\pm(k_x)\right), \\
       \tilde{\mathcal{D}}_\pm(k_x) &= \mathcal{U}_\pm(k_x)  \mathcal{V}^\dagger_\pm(k_x).
    \end{split}
\end{equation}
Putting this together, we find
\begin{equation}
    \nu_-(k_y=0) = +\text{sign} \left(\frac{\Delta}{t_0} \right) = - \nu_+(k_y=0).
\end{equation}
This means that when a mirror-respecting edge is made, at $k_y = 0,$ there is exactly one edge mode belonging to the mirror-odd sector and one edge mode belonging to the mirror-even sector. Then, by chiral symmetry, these modes must be exactly at zero energy since if there was a mode at positive energy in one mirror sector, there must be another one at negative energy in the same mirror sector since chiral symmetry commutes with mirror symmetry.

\subsection{Dirac continuum description of armchair edge states}

The edge states predicted by the topological classification can be obtained directly in the continuum limit. Let us assume that $|\Delta|, |\mu| \ll |t_0|$ so that the relevant physics lies at the Dirac cones of monolayer graphene. Then, we can expand the BdG Hamiltonian around $K$ and $K'$ to find 
\begin{equation}
\begin{split}
    \mathcal{H}^\text{BdG}(\mathbf{k}) &= \begin{pmatrix}
    \mathcal{H}_{K}(\mathbf{k}) & 0 \\
    0 & \mathcal{H}_{K'}(\mathbf{k})
    \end{pmatrix}, \\
    \mathcal{H}_{K}(\mathbf{k}) & = \begin{pmatrix}
    \mu & \hbar v_F \left( k_x - i k_y \right) & -\frac{3 \sqrt{3}}{2}  \Delta & 0 \\
    \hbar v_F \left( k_x + i k_y \right) & \mu & 0 &  -\frac{3 \sqrt{3}}{2} \Delta \\
    -\frac{3 \sqrt{3}}{2}  \Delta & 0 & -\mu & -\hbar v_F \left( k_x - i k_y \right) \\
    0 & -\frac{3 \sqrt{3}}{2}  \Delta  & -\hbar v_F \left( k_x + i k_y \right) & -\mu 
    \end{pmatrix} , \\
    \mathcal{H}_{K'}(\mathbf{k}) & = \begin{pmatrix}
    \mu & \hbar v_F \left( -k_x - i k_y \right) & +\frac{3 \sqrt{3}}{2}  \Delta & 0 \\
    \hbar v_F \left( -k_x + i k_y \right) & \mu & 0 &  +\frac{3 \sqrt{3}}{2} \Delta \\
    +\frac{3 \sqrt{3}}{2}  \Delta & 0 & -\mu & -\hbar v_F \left( -k_x - i k_y \right) \\
    0 & +\frac{3 \sqrt{3}}{2}  \Delta  & -\hbar v_F \left( -k_x + i k_y \right) & -\mu 
    \end{pmatrix}.
\end{split}
\end{equation}
We are interested in solving the wavefunction in real space where $k_\mu = -i \partial_{\mu}.$ The basis state is labeled as
\begin{equation}
    \psi(\mathbf{r}) = \begin{pmatrix} f_{A,K}(\mathbf{r}) & f_{B,K}(\mathbf{r}) & g_{A,K}(\mathbf{r}) & g_{B,K}(\mathbf{r}) & f_{A,K'}(\mathbf{r}) & f_{B,K'}(\mathbf{r}) & g_{A,K'}(\mathbf{r}) & g_{B,K'}(\mathbf{r}) \end{pmatrix}^T.
\end{equation}
We are interested in solving the wavefunction at an armchair edge which we locate at $x = 0$ and the sample extending in the $x\rightarrow - \infty $ direction. We take the boundary condition to be
\begin{equation}
\label{eq: BC 1}
    f_{\sigma,K}(x = 0) = f_{\sigma,K'}(x = 0), \quad g_{\sigma,K}(x = 0) = g_{\sigma,K'}(x = 0).
\end{equation}
Actually, the correct physically-derived boundary condition that requires both sublattices to vanish at the interface should be 
\begin{equation}
\label{eq: BC 2}
    f_{\sigma,K}(x = 0) = -f_{\sigma,K'}(x = 0), \quad g_{\sigma,K}(x = 0) = -g_{\sigma,K'}(x = 0).
\end{equation}
However, Eq. \eqref{eq: BC 1} and Eq. \eqref{eq: BC 2} are related to each other by a simple transformation of the basis functions. So we will continue to use Eq. \eqref{eq: BC 1}. If necessary, one can multiply the spinor in the $K'$ valley by $-1.$ With the boundary condition of Eq. \eqref{eq: BC 1}, instead of solving the wavefunction in the half-plane, we can extend the domain to the entire plane where the wavefunction on the left $(x<0)$ belongs to the $K$ valley while the wavefunction on the right belongs to the $K'$ valley. This amounts to mapping $x \mapsto -x$ on the right half-plane. The boundary condition is satisfied automatically if the wavefunction is continuous at $x = 0.$ This trick was used in Ref. \cite{Abanin2006Spin}. However, $\Delta$ is now spatially dependent
\begin{equation}
    \zeta(x) = \begin{cases} 
      -\frac{3\sqrt{3}}{2}\Delta & x\leq 0 \\
      +\frac{3\sqrt{3}}{2}\Delta & x > 0
   \end{cases}.
\end{equation}
First, let us focus on the zero-energy modes that occur at $k_y = 0.$ Then going into the mirror basis, we obtain
\begin{equation}
    \mathcal{H}(\mathbf{r}) = \left(
\begin{array}{cccc}
 -\mu -i \hbar v_F\partial_x   & \zeta(x)  & 0 & 0 \\
 \zeta(x) & \mu +i \hbar v_F \partial_x  & 0 & 0 \\
 0 & 0 & -\mu +i \hbar v_F\partial_x   & \zeta(x)  \\
 0 & 0 & \zeta(x)  & \mu -i \hbar v_F\partial_x   \\
\end{array}
\right).
\end{equation}
The upper block belongs to the mirror-odd sector, and the bottom block belongs to the mirror-even sector. The problem now is essentially just solving a $2\times2$ Hamiltonian. There are two solutions: one belonging to each mirror sector. If $\Delta > 0,$ then the normalizable, but non-normalized, solutions are
\begin{equation}
    \psi_{m_y=-1}(\mathbf{r}) = \exp \left(\frac{-\zeta(x) +i \mu }{\hbar v_F}x \right) \begin{pmatrix}
    i \\
    1 \\ 
    0 \\
    0
    \end{pmatrix}, \quad \psi_{m_y=+1}(\mathbf{r}) = \exp \left(\frac{-\zeta(x) -i \mu }{\hbar v_F}x \right) \begin{pmatrix}
    0 \\
    0 \\ 
    -i \\
    1
    \end{pmatrix}.
\end{equation}
If $\Delta < 0,$ then the normalizable, but   non-normalized, solutions are
\begin{equation}
    \psi_{m_y=-1}(\mathbf{r}) = \exp \left(\frac{\zeta(x) + i \mu }{\hbar v_F}x \right) \begin{pmatrix}
    -i \\
    1 \\ 
    0 \\
    0
    \end{pmatrix}, \quad \psi_{m_y=+1}(\mathbf{r}) = \exp \left(\frac{\zeta(x) -i \mu }{\hbar v_F}x \right) \begin{pmatrix}
    0 \\
    0 \\ 
    i \\
    1
    \end{pmatrix}.
\end{equation}
We can actually find the states away from zero energy which are connected to these zero-energy states using lowest-order degenerate perturbation theory. They are
\begin{equation}
\begin{split}
   \mathcal{E}(k_y) = \pm \hbar v_F k_y \frac{1}{1+\left(\frac{2\mu}{3\sqrt{3} \Delta}\right)^2}: \quad  \psi_{\mathcal{E} = \pm }(\mathbf{r}) = e^{i k_yy} \left[\pm i \psi_{m_y=-1}(\mathbf{r})  + \psi_{m_y=+1}(\mathbf{r})  \right].
\end{split}
\end{equation}

Next, we consider how to spatially isolate a single Majorana mode. To do this, we add a perturbation along the $y$ direction that breaks mirror symmetry. One such perturbation is a $\sigma_z$ mass (in the original sublattice basis) that changes sign
\begin{equation}
    \lim_{y \rightarrow \pm \infty}m(y) = \pm m,
\end{equation}
where $2m$ denotes the energy difference between the $A$ and $B$ sublattices. Strictly speaking, we should only add the perturbation at the edge because we want to preserve the topological classification of the bulk. In this case, we should multiply the mass term by a Dirac delta function $\delta(x)$ to limit the spatial extent of the perturbation.   However, it is analytically simpler to let the perturbation be uniform in $x.$ Since we only care about the boundary states, the effect of the mass deep inside the bulk should be minimal on the edge states. The Hamiltonian is modified to
\begin{equation}
    \mathcal{H}(\mathbf{r}) = \left(
\begin{array}{cccc}
 -i\hbar v_F \partial_x -\mu  & \zeta(x) & -\hbar v_F \partial_y + m(y) & 0 \\
 \zeta(x) & \mu +i\hbar v_F \partial_x  & 0 & \hbar v_F \partial_y - m(y) \\
+\hbar v_F \partial_y + m(y) & 0 & +i\hbar v_F \partial_x -\mu  & \zeta(x) \\
 0 & -\hbar v_F \partial_y - m(y) & \zeta(x) & -i\hbar v_F \partial_x +\mu  \\
\end{array}
\right).
\end{equation}
The two possible zero-energy solutions are (for $\Delta >0$)
\begin{equation}
\begin{split}
    \psi_{1}(\mathbf{r}) &= \exp \left(\frac{-\zeta(x) +i \mu }{\hbar v_F}x \right) \exp \left(- \frac{1}{\hbar v_F}\int_0^y m(y') dy'  \right) \begin{pmatrix}
    i \\
    1 \\ 
    0 \\
    0
    \end{pmatrix},\\
    \psi_{2}(\mathbf{r}) &= \exp \left(\frac{-\zeta(x) -i \mu }{\hbar v_F}x \right)  \exp \left(+ \frac{1}{\hbar v_F}\int_0^y m(y') dy'  \right)\begin{pmatrix}
    0 \\
    0 \\ 
    -i \\
    1
    \end{pmatrix}.
\end{split}
\end{equation}
Only one of these solutions is normalizable depending on whether $m$ is positive or negative. The solutions for $\Delta < 0$ are
\begin{equation}
\begin{split}
    \psi_{1}(\mathbf{r}) &= \exp \left(\frac{\zeta(x) +i \mu }{\hbar v_F}x \right) \exp \left(- \frac{1}{\hbar v_F}\int_0^y m(y') dy'  \right) \begin{pmatrix}
    -i \\
    1 \\ 
    0 \\
    0
    \end{pmatrix},\\
    \psi_{2}(\mathbf{r}) &= \exp \left(\frac{\zeta(x) -i \mu }{\hbar v_F}x \right)  \exp \left(+ \frac{1}{\hbar v_F}\int_0^y m(y') dy'  \right)\begin{pmatrix}
    0 \\
    0 \\ 
    i \\
    1
    \end{pmatrix}.
\end{split}
\end{equation}

\subsection{Numerical simulation of finite flakes}

\begin{figure}
    \centering
    \includegraphics[width=7.0in]{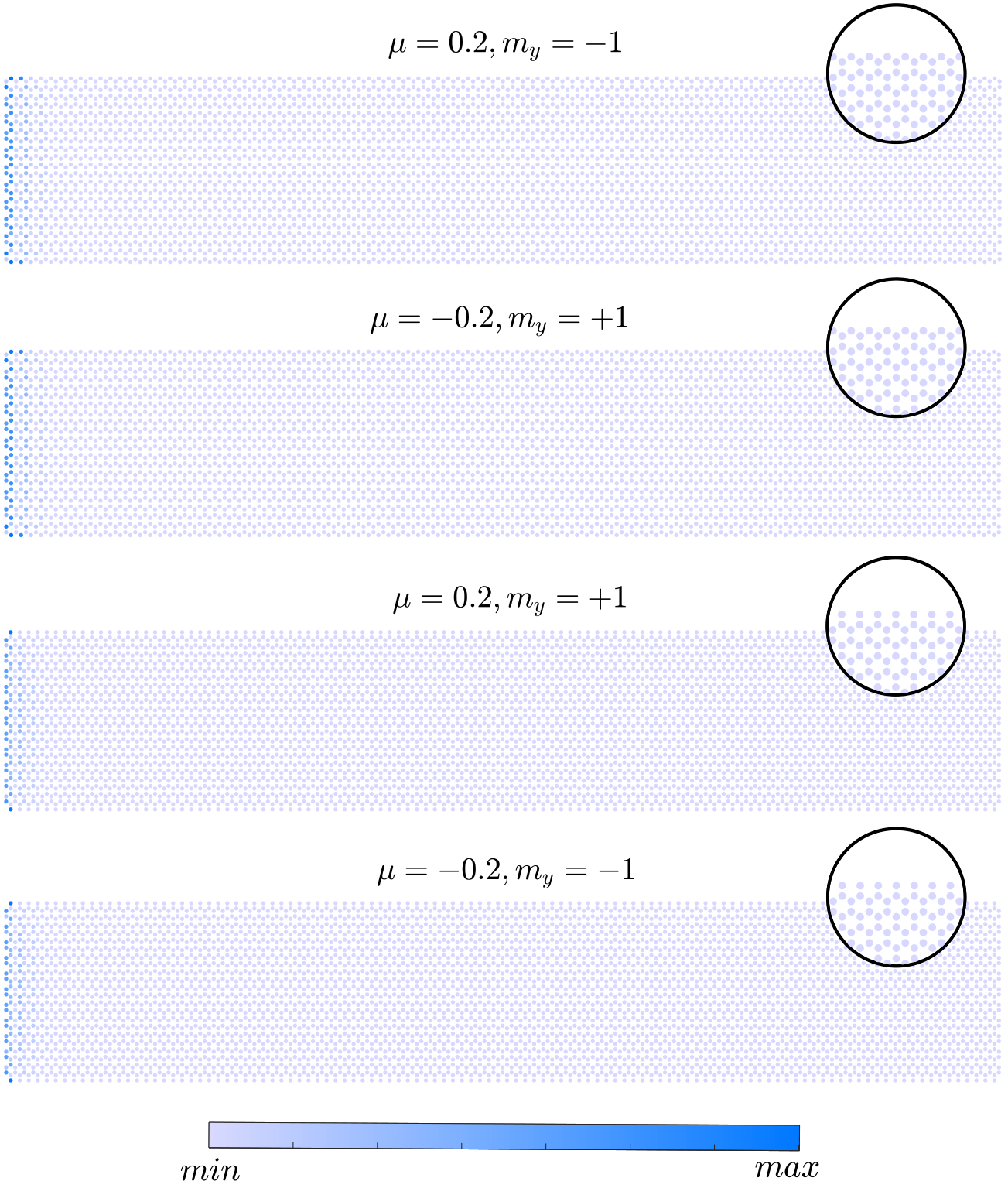}
    \caption{\textbf{Charge density of zero modes in rectangular flakes.} For all simulations, $\Delta = 0.1t_0,$ the lengths are approximately $20a,$ and the widths are approximately $100a-110a.$ Each plot shows a zero-energy eigenstate  localized on the left (the complementary right-localized state does exist, but is not shown). The mirror eigenvalue of the displayed state is shown on top of each plot. We find that as the chemical potential switches sign, the mirror eigenvalue of the zero modes also changes. Here, we only show the charge density of the electron sector. The two upper plots feature zigzag top and bottom edges, and the two lower plots feature bearded top and bottom edges. All energies are approximately $10^{-16}t_0.$ }
    \label{fig:flake1}
\end{figure}

In this section, we numerically diagonalize the Hamiltonian for finite flakes. We first study rectangular flakes with armchair edges along the $y$-axis and zigzag or bearded edges along the $x$-axis. We intentionally make the flakes respect $\mathcal{M}_y$ symmetry so that we can label states by their mirror eigenvalue. Each such flake features only \textit{two} zero-energy states (these states are not \textit{exactly} at zero energy due to finite-size mixing effects, but they have very small energies $\lesssim 10^{-15}$) instead of four zero-energy states expected in the infinite nanoribbon case. This difference in the number of modes is probably due to the imposition of nontrivial boundary conditions at bearded or zigzag edges. Depending on whether the top and bottom edges are zigzag or bearded and whether the chemical potential is positive or negative, the edge states have either positive or negative mirror eigenvalue, as shown in Fig. \ref{fig:flake1}. In other words, the presence of zigzag or bearded edges selects out one of the two zero modes per edge with definite mirror character, and completely suppresses the other zero mode. This behavior is very similar to the effect of a $\sigma_z$ mass. Therefore, under appropriate conditions, one can heuristically think of the imposition of zigzag and bearded edges as having a varying $\sigma_z$ mass that is arbitrarily large (with the appropriate signs) outside the sample and zero inside the sample. We have only inferred this connection, and have \textit{not} rigorously shown it to be true, but it seems to explain the number of modes as well as their mirror character obtained numerically.

\begin{figure}
    \centering
    \includegraphics[width=7.0in]{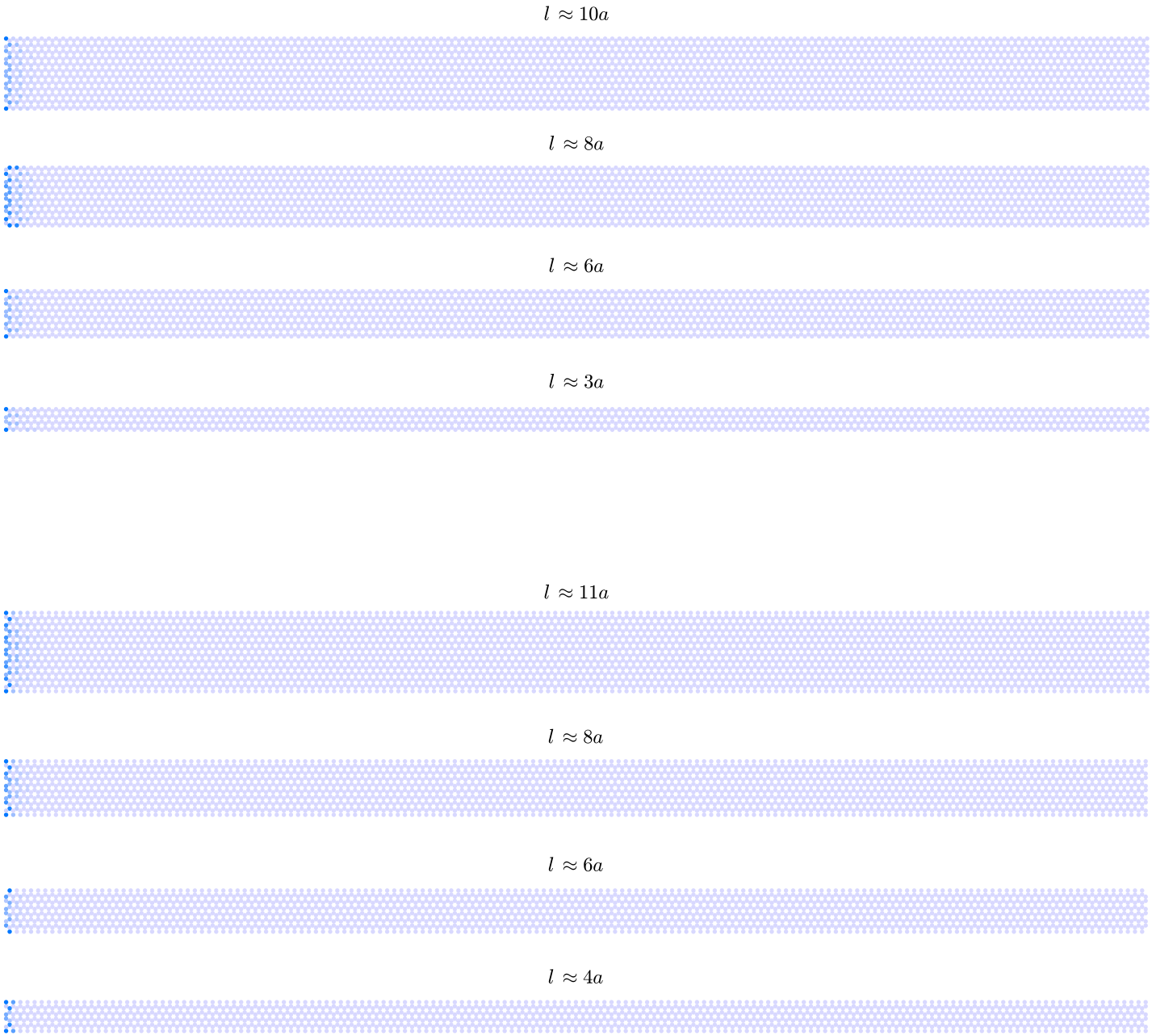}
    \caption{\textbf{Zero modes as a function of flake size.} For all simulations, $\Delta = 0.1t_0,$ $\mu = 0.2t_0,$ and the  widths are approximately $160a.$ As we vary the length of the flakes, we  observe that the zero modes remain and are localized to the edge. The upper plots feature zigzag edges at the top and bottom, while the lower plots feature bearded edges at the top and  bottom. All energies are approximately $10^{-15}t_0.$  }
    \label{fig:flake2}
\end{figure}

\begin{figure}
    \centering
    \includegraphics[width=7.0in]{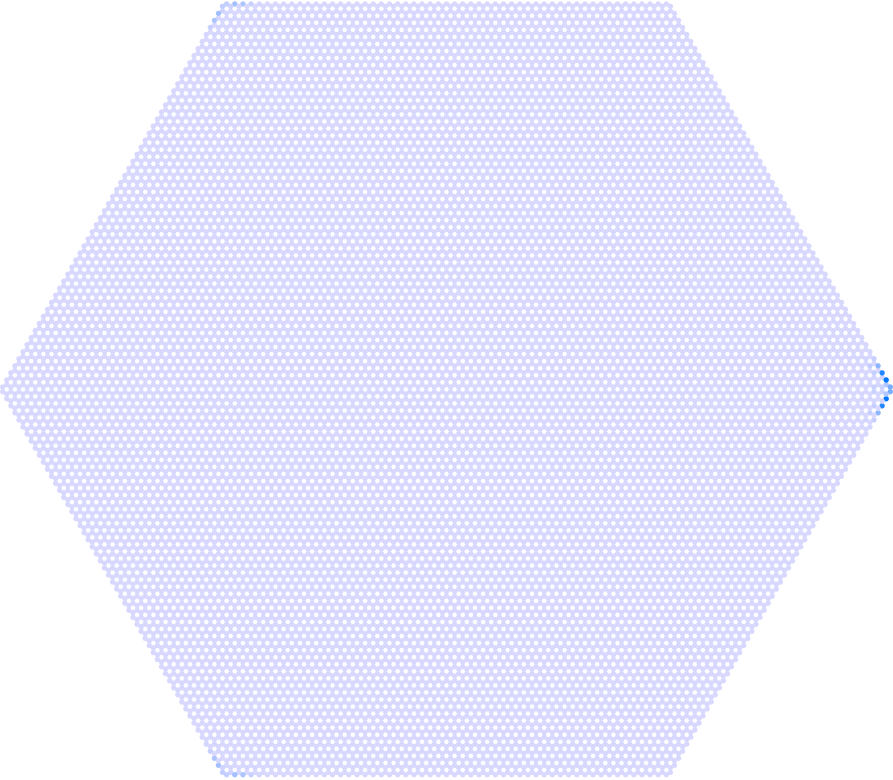}
    \caption{\textbf{Corner zero mode in a hexagonal flake.} A zero mode is trapped at a corner of a hexagonal flake. There are six such zero modes in total.  Here, $\mu = 0.2t_0$ and $\Delta = 0.1t_0.$ The energy is about $10^{-12}t_0.$ }
    \label{fig:flake3}
\end{figure}

\begin{figure}
    \centering
    \includegraphics[width = 6in]{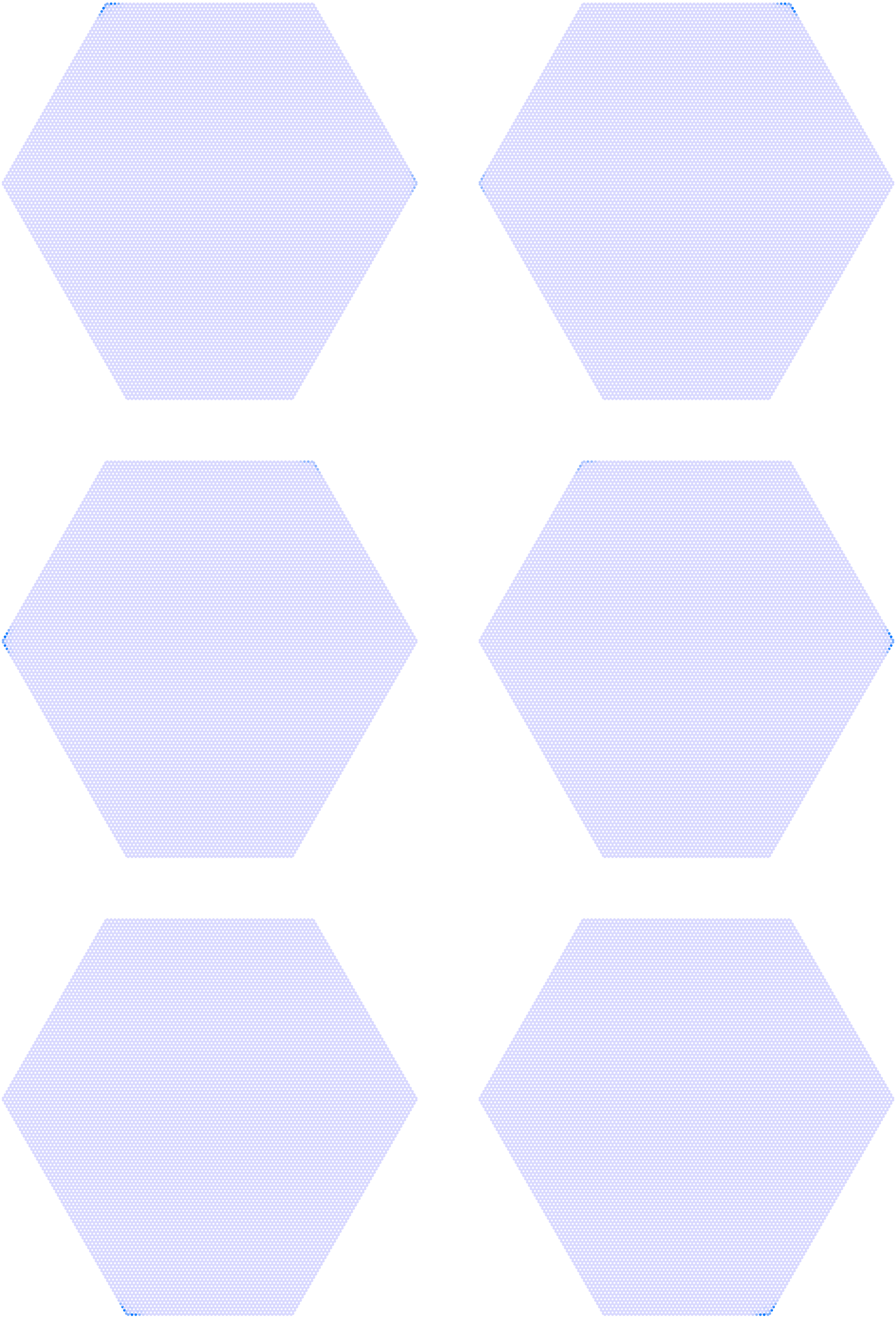}
    \caption{\textbf{Corner modes in a distorted hexagonal flake.} The flake is not a regular hexagon, but instead have non-congruent sides. The energies are $\lesssim 10^{-9} t_0.$ There are six near zero-modes located at the corners of the flake. Here, $\mu = 0.2t_0$ and $\Delta = 0.1t_0.$}
    \label{fig:distorted hexagonal flake}
\end{figure}

To separate the two zero modes spatially on the left and the right of the sample, we need to make the width along the $x$-direction large enough to prevent wavefunction overlap. However, the length along the $y$-direction does not qualitatively affect the presence of the zero modes. In fact, we can make the length microscopic, and the zero modes would still be there. This is true for both zigzag and bearded edges, as shown in Fig. \ref{fig:flake2}. We also simulate a regular hexagonal flake. In this case, we find six zero modes, each localized at a corner of the flake. An example is shown in Fig. \ref{fig:flake3}. The same conclusion holds when the hexagonal flake is \textit{not} regular, as shown in Fig. \ref{fig:distorted hexagonal flake}. All of these finite-size calculations show that the present platform offers a myriad of ways to manipulate the Majorana zero modes by design of appropriate edges and corners.

\subsection{Spin domain walls}

\begin{figure}
    \centering
        \includegraphics[width=5.5in]{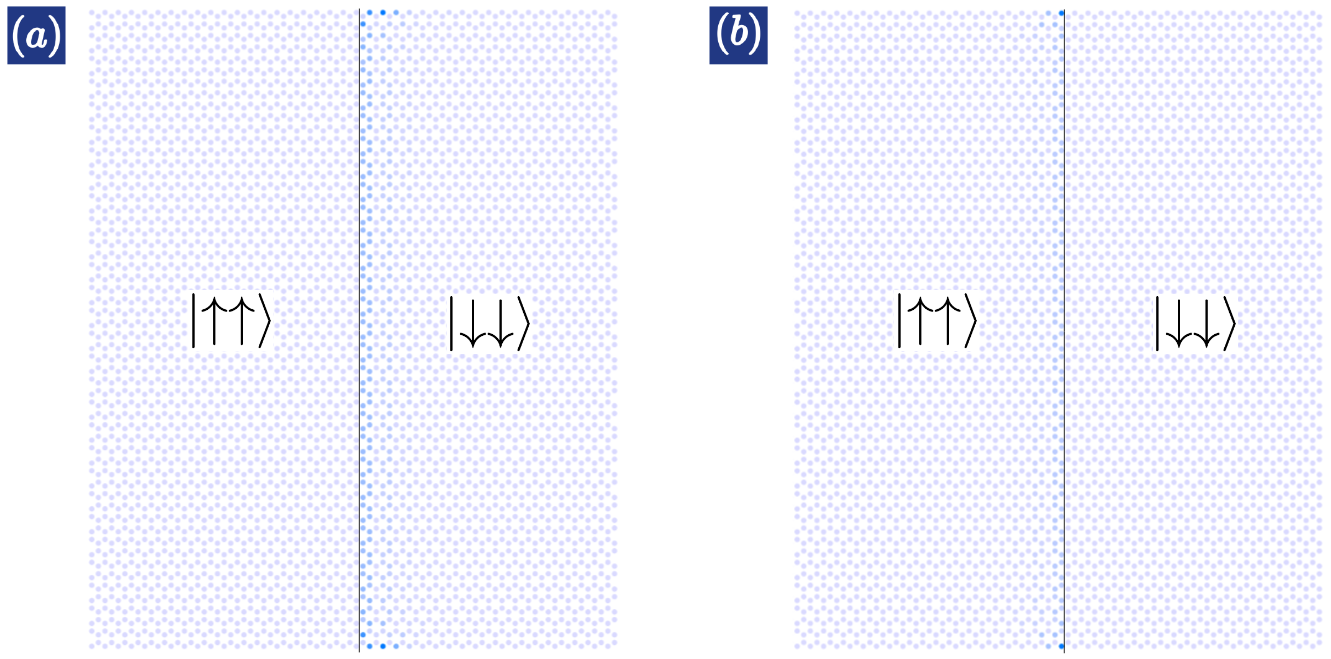}
    \caption{\textbf{Zero modes near a spin domain wall}. Charge maps of two zero modes located near a domain wall between $\ket{\uparrow\uparrow}$ and $\ket{\downarrow\downarrow}$ spin domains. The domain wall is aligned with the armchair direction. Panel (a) shows the zero mode living in the $\ket{\downarrow\downarrow}$ domain, and (b) the other zero mode residing in the $\ket{\uparrow\uparrow}$ domain. The black line indicates the position of the domain wall. The states have energy $E\approx10^{-17}t_0$. In the simulations, $\Delta_\text{sc}\approx0.25t_0$, $\mu\approx0.1t_0,$ and the length of each domain is $l\approx150a$. }
    \label{fig:MZMdomainwall}
\end{figure}

In this paper, we focus on spin-polarized superconductivity. However, spin domain walls can develop in a zero-field-cooled experiment. Here, we point out one situation in which this type of defect can lead to zero modes. Let us suppose that one side hosts $\ket{\uparrow\uparrow}$ Cooper pairs and the other side hosts $\ket{\downarrow\downarrow}$ Cooper pairs. Consequently, absent spin-flip processes, these Cooper pairs cannot tunnel between the domain wall. Then, from either side of the wall, quasiparticles see a superconductor/vacuum interface, so each domain behaves effectively as an independent system. Therefore, in a finite flake with two such domains, if mirror symmetry is respected within each domain, and the domain wall is aligned along an armchair crystal axis, then there are \textit{two} zero modes at the domain wall, one coming from each domain, see Fig. \ref{fig:MZMdomainwall}. It is worth noting that these states do not exist if the pairing is $s$-wave. So domain walls can also serve as a diagnostic for $f$-wave superconductivity.

\section{Rhombohedral multilayer graphene}

In this section, we generalize the previous results to multilayer systems.

\subsection{Bernal bilayer graphene}

\begin{figure}
    \centering
    \includegraphics[width=5in]{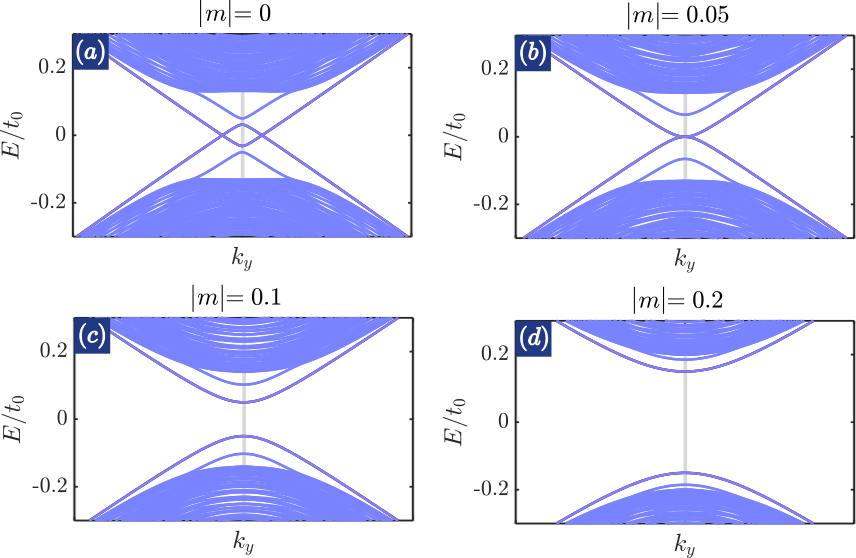}
    \caption{\textbf{Band structure of superconducting Bernal bilayer graphene.} (a) There are \textit{no} zero modes at $k_y = 0,$ but there are zero modes away from $k_y = 0.$ (b)-(d) By adding a mass term of the form Eq. \eqref{eq: bilayer mass}, these non-topological zero modes can be eliminated without breaking $\mathcal{M}_y$ or closing the bulk gap. Here, $\mu = \Delta = 0.05t_0.$ There is no qualitative difference between positive and negative values of $m.$}
    \label{fig:bilayer_band_structure}
\end{figure}

The Hamiltonian for Bernal bilayer graphene is just that of two copies of monolayer graphene with an interlayer coupling $\gamma_1 = 0.1t_0$
\begin{equation}
\mathcal{H}^\text{BdG}(\mathbf{k}) = \begin{pmatrix}
\mu & -t_0 h_\mathbf{k} & 0 & 0 & \Delta g_{\mathbf{k}} & 0 & 0 & 0 \\
-t_0 h^*_\mathbf{k} & \mu & \gamma_1 & 0 & 0 & \Delta g_{\mathbf{k}} & 0 & 0 \\ 
0 & \gamma_1 & \mu & -t_0 h_\mathbf{k} & 0 & 0 & \Delta g_{\mathbf{k}} & 0   \\
0 & 0 & -t_0 h^*_\mathbf{k} & \mu & 0 & 0 & 0 & \Delta g_{\mathbf{k}}  \\
\Delta g_{\mathbf{k}} & 0 & 0 & 0 & -\mu & t_0 h_\mathbf{k} & 0 & 0 \\
0 & \Delta g_{\mathbf{k}} & 0 & 0 & t_0 h^*_\mathbf{k} & -\mu & - \gamma_1 & 0 \\
0 & 0 & \Delta g_{\mathbf{k}} & 0 & 0 & -\gamma_1 & -\mu & t_0 h_\mathbf{k} \\
0 & 0 & 0 & \Delta g_{\mathbf{k}} & 0 & 0 & t_0 h^*_\mathbf{k} & -\mu
\end{pmatrix}.
\end{equation}
Particle-hole, time reversal, and chiral  symmetries have the same representation as before
\begin{equation}
\begin{split}
    \mathcal{P}: &\quad \left(\tau_x\otimes \lambda_0 \otimes \sigma_0 \right)\left[\mathcal{H}^\text{BdG}(\mathbf{k})\right]^*\left(\tau_x\otimes \lambda_0\otimes \sigma_0 \right) = -\mathcal{H}^\text{BdG}(-\mathbf{k}), \\
    \mathcal{T}: &\quad \left(\tau_z\otimes \lambda_0\otimes \sigma_0 \right)\left[\mathcal{H}^\text{BdG}(\mathbf{k})\right]^*\left(\tau_z\otimes \lambda_0\otimes \sigma_0 \right) = \mathcal{H}^\text{BdG}(-\mathbf{k}), \\
    \mathcal{C}: &\quad \left(i\tau_y\otimes \lambda_0 \otimes \sigma_0 \right)\mathcal{H}^\text{BdG}(\mathbf{k})\left(-i\tau_y\otimes \lambda_0 \otimes \sigma_0 \right) = -\mathcal{H}^\text{BdG}(\mathbf{k}),
\end{split}
\end{equation}
where the $\lambda$ Pauli matrices act on the layer degree of freedom. However, mirror symmetry needs a slight modification. Actually, in the context of Bernal bilayer graphene, mirror symmetry is a $\mathcal{C}_{2x}$ rotation symmetry that exchanges the two layers as well as sublattices. Despite that, we continue to use $\mathcal{M}_y$ when we mean $\mathcal{C}_{2x}.$ The mirror symmetry operator is 
\begin{equation}
    \mathcal{M}_y: \quad \left(\tau_0 \otimes \lambda_x \otimes \sigma_x \right) \mathcal{H}^\text{BdG}(\mathbf{k}) \left(\tau_0 \otimes \lambda_x \otimes \sigma_x \right) = \mathcal{H}^\text{BdG}(\mathcal{M}_y\mathbf{k}).
\end{equation}
Along the mirror-symmetric line along $k_y=0$, we can again block-diagonalize the Hamiltonian
\begin{equation}
    \begin{split}
        \mathcal{H}^\text{BdG}(k_x) &= \begin{pmatrix}
        \mathcal{H}_-(k_x) & 0 \\
        0 & \mathcal{H}_+(k_x)
        \end{pmatrix}, 
    \end{split}
\end{equation}
where
\begin{equation}
    \begin{split}
        \mathcal{H}_-(k_x) & = \left(
\begin{array}{cccc}
 -\mu  & -h(k_x) & g(k_x) & 0 \\
 -h(k_x) & \gamma_1 -\mu  & 0 & g(k_x) \\
 g(k_x) & 0 & \mu  & h(k_x) \\
 0 & g(k_x) & h(k_x) & \mu -\gamma_1  \\
\end{array}
\right), \\
\mathcal{H}_+(k_x) &= \left(
\begin{array}{cccc}
 -\mu  & -h(k_x) & g(k_x) & 0 \\
 -h(k_x) & -\gamma_1 -\mu  & 0 & g(k_x) \\
 g(k_x) & 0 & \mu  & h(k_x) \\
 0 & g(k_x) & h(k_x) & \gamma_1 +\mu  \\
\end{array}
\right).
    \end{split}
\end{equation}
Putting into chiral form, we obtain
\begin{equation}
\begin{split}
    \mathcal{H}_\pm(k_x) &= \begin{pmatrix}
    0 & \mathcal{D}_\pm(k_x) \\
    \mathcal{D}^\dagger_\pm(k_x) & 0
    \end{pmatrix}, \\
    \mathcal{D}_\pm(k_x) &= \left(
\begin{array}{cc}
 \pm \gamma_1 -i g(k_x)+\mu  & h(k_x) \\
 h(k_x) & \mu -i g(k_x) \\
\end{array}
\right).
\end{split}
\end{equation}
Calculating the winding number, we find that $\nu_\pm = 0$ for much of parameter space. So, this system is topologically trivial. This is confirmed by band structure calculation on an armchair nanoribbon that respects $\mathcal{M}_y$ symmetry, shown in Fig. \ref{fig:bilayer_band_structure}(a). There, we see no zero modes at $k_y = 0.$ However, interestingly, we do see zero modes displaced away from $k_y = 0$ for the present model. These are ascertained numerically to a precision of $ \sim 10^{-15}.$ So, as far as we can tell numerically, they are at exact zero energy. These are not topological though, as is clear from the fact that they are connected to the same bulk bands. To confirm this, we add a layer-antisymmetric $\sigma_z$ mass of the form 
\begin{equation}
\label{eq: bilayer mass}
    V = m \tau_z \otimes \lambda_z \otimes \sigma_z.
\end{equation}
This potential energy respects $\mathcal{M}_y$ because it simultaneously flips sign under layer and sublattice exchange. So it commutes with $\mathcal{M}_y.$ By increasing $m,$ we can remove the zero modes at $k_y \neq 0,$ as shown in Fig. \ref{fig:bilayer_band_structure}(b)-(d), without closing the bulk gap, illustrating that these modes are accidental.

\subsection{Rhombohedral trilayer graphene}

\begin{figure}
    \centering
    \includegraphics[width=5in]{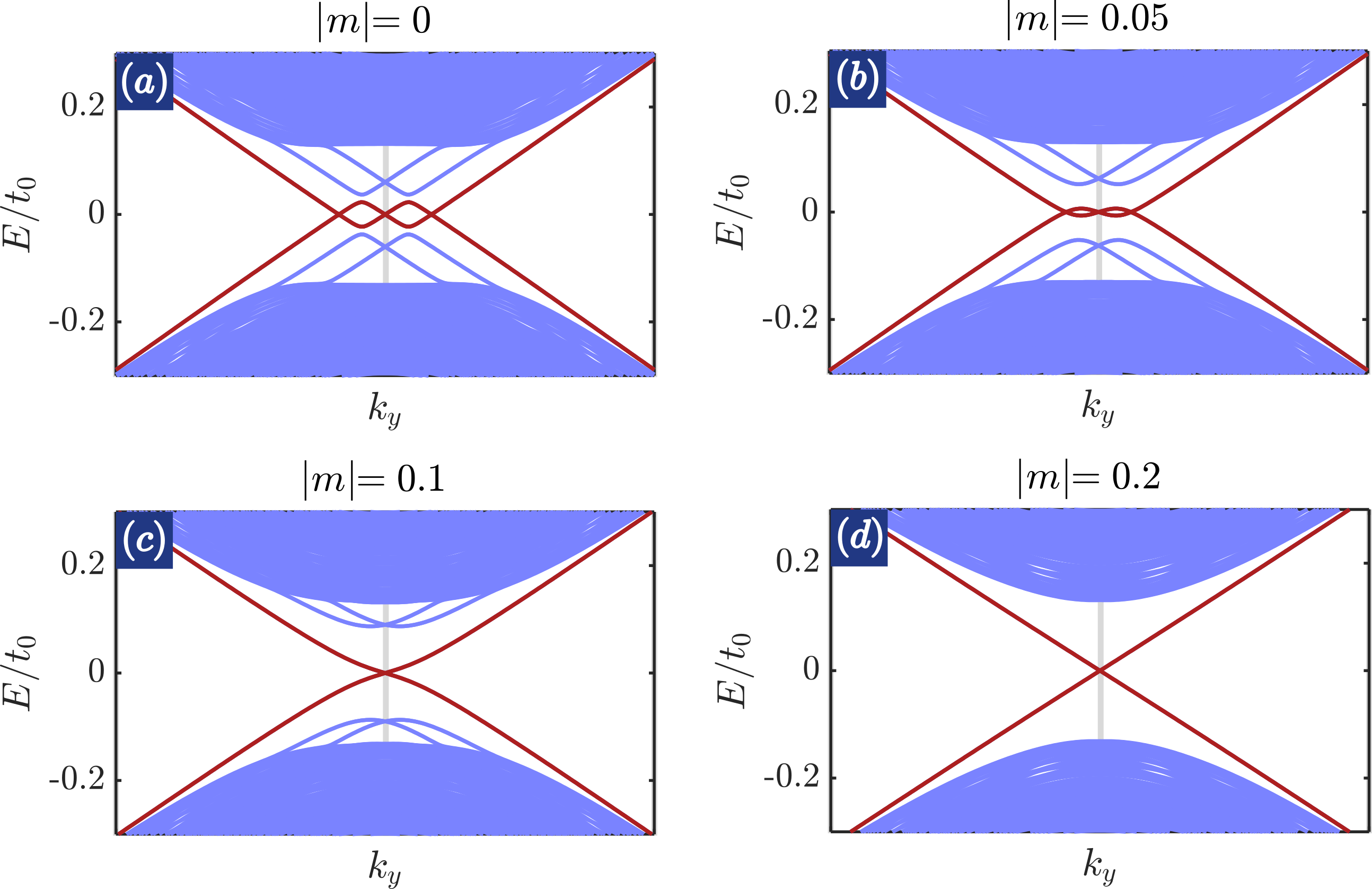}
    \caption{\textbf{Band structure of superconducting rhombohedral trilayer graphene.} (a) There are two zero modes per edge at $k_y = 0;$ there are also zero modes away from $k_y = 0.$ (b)-(d) By adding a mass term of the form Eq. \eqref{eq: mass in trilayer}, these non-topological zero modes can be eliminated without breaking $\mathcal{M}_y$ or closing the bulk gap. However, the topological zero modes at $k_y=0$ are \textit{not} gapped since they are robust. Here, $\mu = \Delta = 0.05t_0.$ There is no qualitative difference between positive and negative values of $m.$ }
    \label{fig:trilayer_band_structure}
\end{figure}

The Hamiltonian for superconducting rhombohedral trilayer graphene is 
\begin{equation}
    \mathcal{H}^\text{BdG}(\mathbf{k}) = \begin{pmatrix}
    \mathcal{H}_3 (\mathbf{k}) & \mathcal{H}_\text{sc}(\mathbf{k}) \\
    \mathcal{H}_\text{sc}^\dagger(\mathbf{k}) & -\mathcal{H}_3 (\mathbf{k})
    \end{pmatrix},
\end{equation}
where 
\begin{equation}
    \begin{split}
        \mathcal{H}_3(\mathbf{k}) &= \begin{pmatrix}1 & 0 & 0\\
        0 & 1 & 0 \\
        0 & 0 & 1
        \end{pmatrix} \otimes \mathcal{H}_\text{intra}(\mathbf{k})  + \begin{pmatrix}0 & 1 & 0\\
        0 & 0 & 1\\
        0 & 0 & 0
        \end{pmatrix} \otimes \mathcal{H}_\text{inter}(\mathbf{k}) + \begin{pmatrix}0 & 0 & 0\\
        1 & 0 & 0\\
        0 & 1 & 0
        \end{pmatrix} \otimes \mathcal{H}^\dagger_\text{inter}(\mathbf{k}),  \\
        \mathcal{H}_\text{intra}(\mathbf{k}) &= \begin{pmatrix}
        \mu & -t_0 h_\mathbf{k} \\
        -t_0 h^*_\mathbf{k} & \mu
        \end{pmatrix}, \\
        \mathcal{H}_\text{inter}(\mathbf{k}) &= \begin{pmatrix}
        0 & 0 \\
        \gamma_1 & 0
        \end{pmatrix}, \\
        \mathcal{H}_\text{sc}(\mathbf{k}) &= \begin{pmatrix}1 & 0 & 0\\
        0 & 1 & 0 \\
        0 & 0 & 1
        \end{pmatrix} \otimes \begin{pmatrix}\Delta g_\mathbf{k} & 0\\
        0 & \Delta g_\mathbf{k}  
        \end{pmatrix}.
    \end{split} 
\end{equation}
The mirror operator is represented by 
\begin{equation}
    \mathcal{M}_y = \tau_0 \otimes \begin{pmatrix} 0 & 0 & 1 \\
    0 & 1 & 0 \\
    1 & 0 & 0 
    \end{pmatrix} \otimes \sigma_x,
\end{equation}
which exchanges the top and bottom layers, but maps the middle layer to itself. We also add a mirror-preserving potential
\begin{equation}
\label{eq: mass in trilayer}
    V = m \tau_z \otimes \begin{pmatrix} 1 & 0 & 0 \\
    0 & 0 & 0 \\
    0 & 0 & -1 
    \end{pmatrix} \otimes \sigma_z.
\end{equation}
The winding number can be calculated as before. We find that $\nu_\pm = \pm 1$ for this system. So this system is topologically nontrivial similar to the monolayer system. We confirm this with finite calculations on an armchair nanoribbon. We find two Majorana zero modes per boundary pinned to $k_y = 0,$ as shown in Fig. \ref{fig:trilayer_band_structure}(a). There are other zero modes at $k_y \neq 0,$ but these are \textit{not} topological. They can be eliminated by $\mathcal{M}_y$-symmetric potentials without closing the bulk gap, as illustrated in Fig. \ref{fig:trilayer_band_structure}(b)-(d).

\begin{figure}
    \centering
    \includegraphics[width=3in]{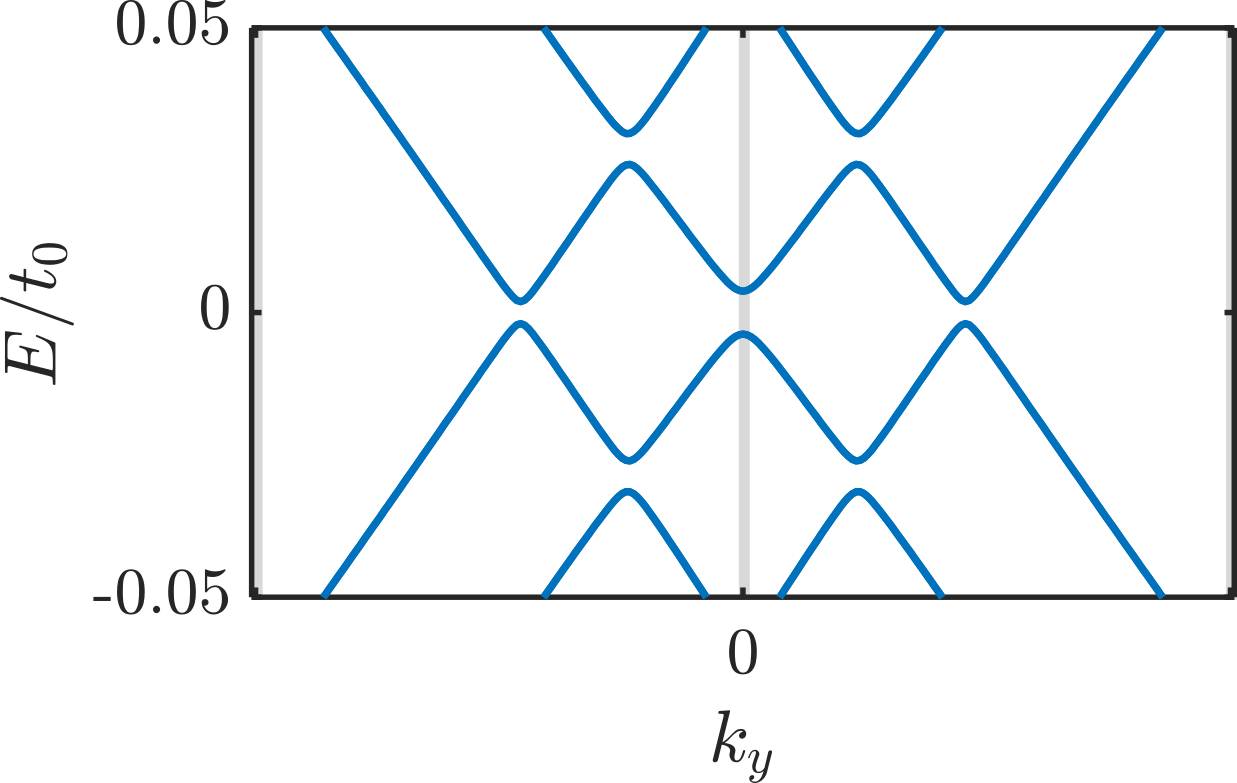}
    \caption{\textbf{Electric-field-induced hybridization of zero modes in rhombohedral trilayer graphene.} With the addition of a perturbation in the form of Eq. \eqref{eq: electric field in trilayer}, the zero modes at $k_y = 0$ are gapped out because of broken mirror symmetry. Here, $\mu = \Delta = 0.1t_0$ and $|V_{0}| = 0.1 t_0.$ }
    \label{fig:electricfield}
\end{figure}

In trilayer graphene and its multilayer generalization, the layer equivalence can be broken simply by adding a perpendicular electric field. The perturbation is described by 
\begin{equation}
\label{eq: electric field in trilayer}
    V = V_0 \tau_z \otimes \begin{pmatrix} 1 & 0 & 0 \\
    0 & 0 & 0 \\
    0 & 0 & -1 
    \end{pmatrix} \otimes \sigma_0.
\end{equation}
This electric field hybridizes the topological zero modes and gap them out, as shown in Fig. \ref{fig:electricfield}.

\subsection{Rhombohedral $N$-layer generalization}

\begin{figure}
    \centering
    \includegraphics[width=5in]{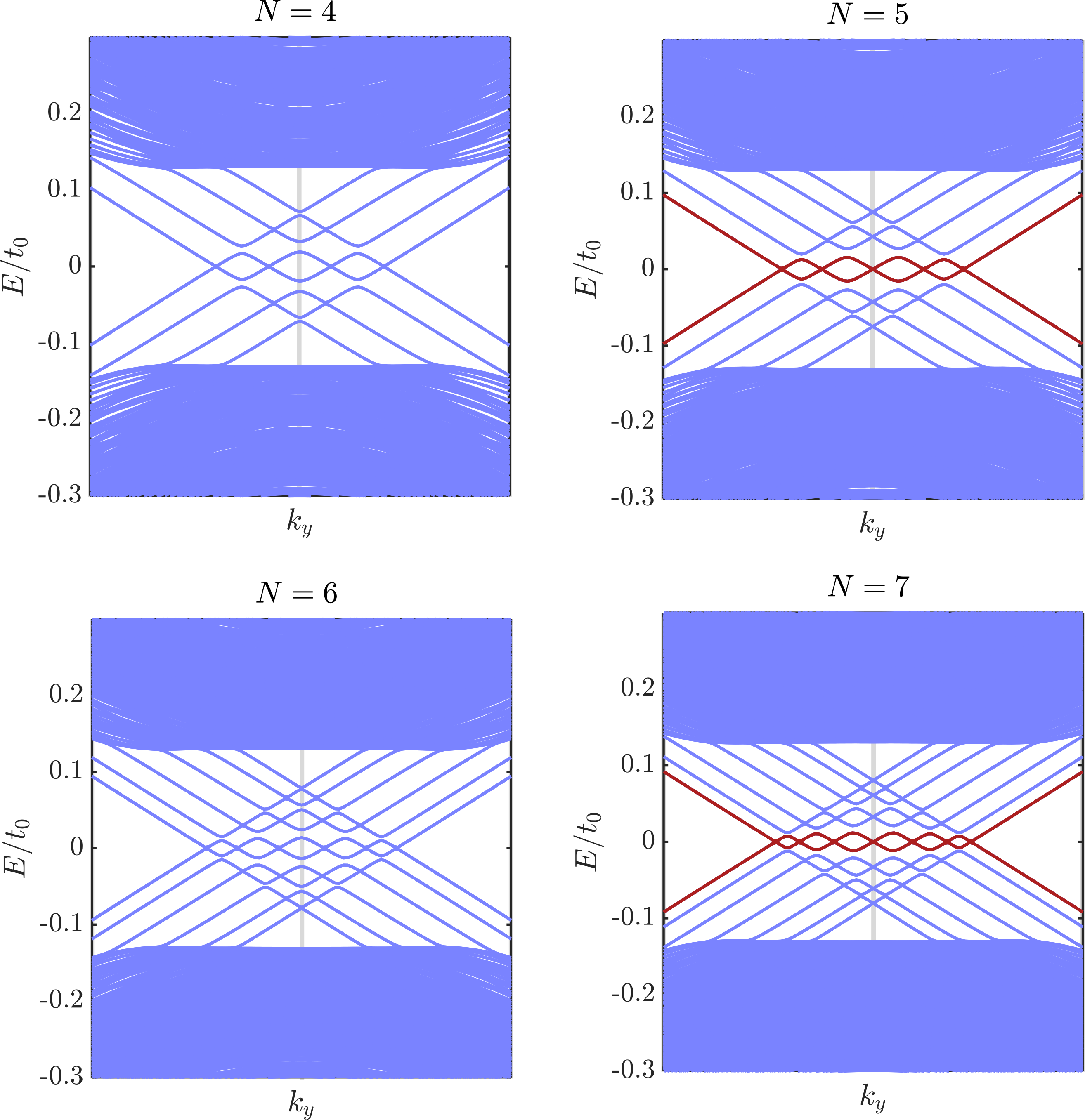}
    \caption{\textbf{Band structure of armchair nanoribbons of multilayer stacks.} We observe that for stacks with an odd number of layers, there are zero modes at $k_y =0,$ while for stacks with an even number of layers, there are not. }
    \label{fig:multiplelayers}
\end{figure}

The generalization to $N>1$ layers is straightforward. The Hamiltonian is
\begin{equation}
    \mathcal{H}^\text{BdG}(\mathbf{k}) = \begin{pmatrix}
    \mathcal{H}_N (\mathbf{k}) & \mathcal{H}_\text{sc}(\mathbf{k}) \\
    \mathcal{H}_\text{sc}^\dagger(\mathbf{k}) & -\mathcal{H}_N (\mathbf{k})
    \end{pmatrix},
\end{equation}
where 
\begin{equation}
    \begin{split}
        \mathcal{H}_N(\mathbf{k}) &= \mathbb{I}_{N \times N} \otimes \mathcal{H}_\text{intra}(\mathbf{k})  + \mathbb{U}_{N \times N}  \otimes \mathcal{H}_\text{inter}(\mathbf{k}) + \mathbb{L}_{N \times N}  \otimes \mathcal{H}^\dagger_\text{inter}(\mathbf{k}),  \\
        \mathcal{H}_\text{intra}(\mathbf{k}) &= \begin{pmatrix}
        \mu & -t_0 h_\mathbf{k} \\
        -t_0 h^*_\mathbf{k} & \mu
        \end{pmatrix}, \\
        \mathcal{H}_\text{inter}(\mathbf{k}) &= \begin{pmatrix}
        0 & 0 \\
        \gamma_1 & 0
        \end{pmatrix}, \\
        \mathcal{H}_\text{sc}(\mathbf{k}) &= \mathbb{I}_{N \times N}  \otimes \begin{pmatrix}\Delta g_\mathbf{k} & 0\\
        0 & \Delta g_\mathbf{k}  
        \end{pmatrix}.
    \end{split} 
\end{equation}
The mirror operator is represented by 
\begin{equation}
    \mathcal{M}_y = \tau_0 \otimes \bar{\mathbb{I}}_{N \times N} \otimes \sigma_x.
\end{equation}
Here, $\mathbb{I}_{N \times N} $ is the $N\times N$ identity matrix, $\mathbb{U}_{N \times N} $ ($\mathbb{L}_{N \times N} $) is the sparse $N\times N$ matrix with ones along the upper (lower) diagonal right above (below) the principal diagonal, and $\bar{\mathbb{I}}_{N \times N}$ is the anti-diagonal matrix of ones. We note that for $N$ even, all layers are flipped under mirror symmetry, while for $N$ odd, there is one central layer which is mapped to itself under mirror symmetry. Carrying out the same topological classification exactly as before, we find that the even-$N$ stacks are topologically trivial, while the odd-$N$ stacks are topologically nontrivial protected by $\mathcal{M}_y$ symmetry with winding numbers $\nu_\pm = \pm 1.$ We have checked this fact numerically for $N \leq 50.$ Though we have not proven this trend to be true for all $N,$ it seems likely that it is true for all values of $N.$ The armchair band structures for a few values of $N$ are shown in Fig. \ref{fig:multiplelayers}.

\section{Vortex-confined zero modes}

\begin{figure}
    \centering
    \includegraphics[width=6in]{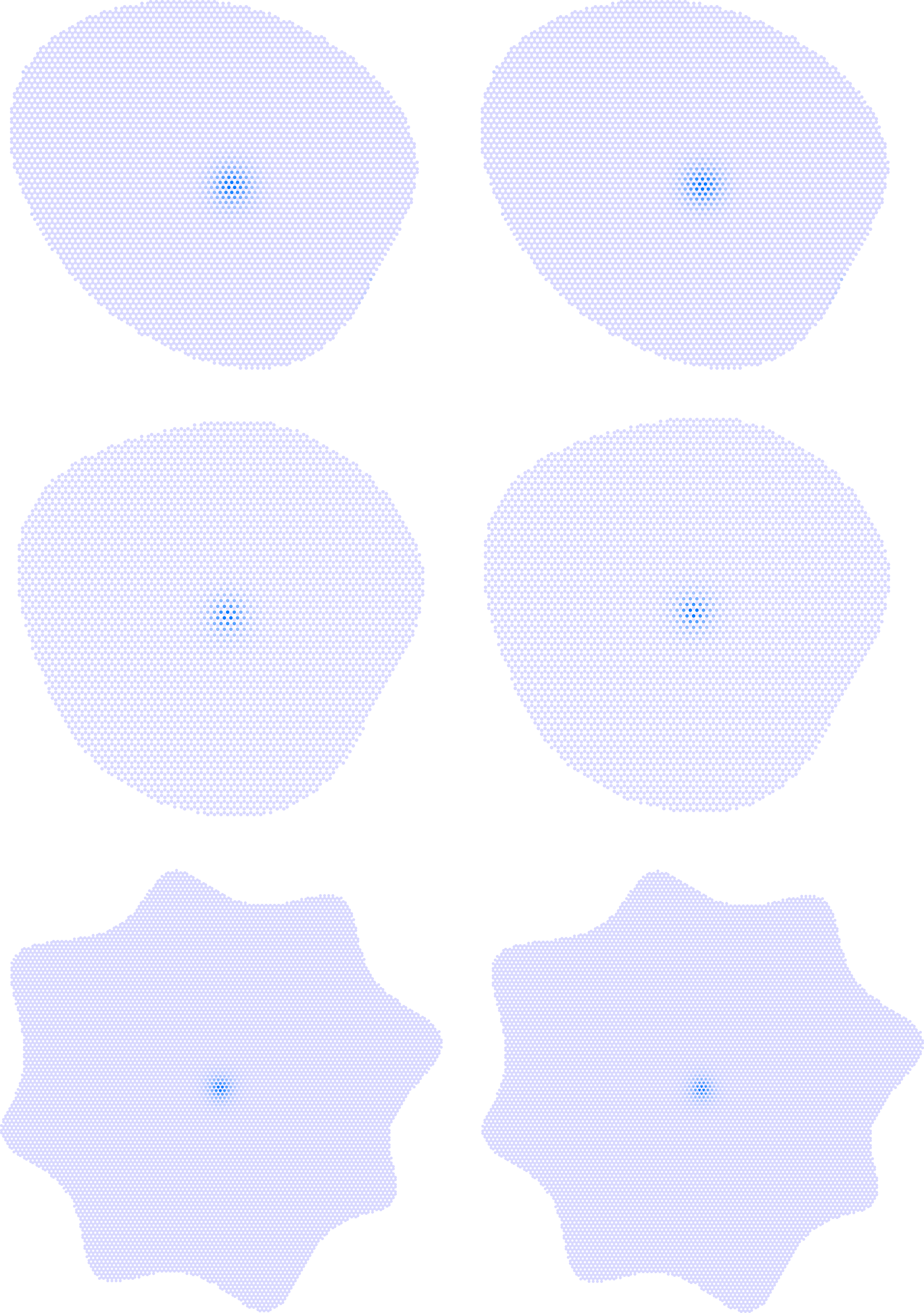}
    \caption{\textbf{Vortex states in monolayer graphene.}  Each flake has two modes near zero energy trapped at the vortex, which is located at the center. The flakes are intentionally irregularly shaped to illustrate that the vortex states are \textit{not} sensitive to boundary conditions. Here, $\mathbf{r}  =\mathbf{0}$ is located at the center of a hexagon. We use $\Delta = \mu = 0.1 t_0.$}
    \label{fig:monolayervortex}
\end{figure}

In this section, we study zero modes confined to vortex cores numerically in the tight-binding formalism. To do this, we only change the next-nearest-neighbor superconducting hopping parameters. Without a vortex, the hopping $\Delta(\mathbf{r}_i, \mathbf{r}_j) = \pm\Delta/2i$ is a constant for $|\mathbf{r}_i - \mathbf{r}_j| = a$ and zero otherwise. With a vortex located at $\mathbf{r} = \mathbf{0},$ we simply modify this hopping to include a spatial dependence
\begin{equation}
    \Delta(\mathbf{r}_i, \mathbf{r}_j) = \pm\frac{ \Delta}{2i} e^{ \pm  i\phi_\mathbf{r}} \tanh \left( \frac{|\mathbf{r}|}{d} \right),
\end{equation}
where $d$ is the distance over which the superconducting gap vanishes,  $\mathbf{r} = (\mathbf{r}_i + \mathbf{r}_j)/2$ is the center-of-mass location, and  $\phi_\mathbf{r}$ is the angle of that position. As required, the superconducting order parameter vanishes exactly at the vortex $\mathbf{r} = \mathbf{0}.$ For monolayer graphene, we find that there are two zero modes trapped at each vortex, as shown in Fig. \ref{fig:monolayervortex}, when the center of the vortex is at the center of a hexagon or a mid-bond or another location along a $\mathcal{M}_y$-symmetric line. When the position of the vortex center is at a random location, then these low-energy modes are not exactly at zero. However, when $d \gtrsim a,$ then the precise center of the vortex is not as important, so the low-energy states are practically at zero energy for various combinations of parameters. We have checked that the existence of the vortex low-energy pairs is not sensitive to the location of the center $\mathbf{r} = \mathbf{0}$ when $d \gtrsim a.$ In Fig. \ref{fig:monolayervortex}, we intentionally choose shapes with irregular boundaries to show that the vortex states are not sensitive to boundary conditions when the flakes are large.

\section{$ABA$ multilayer graphene}

In this section, we study $ABA$ multilayer graphene, which is the most stable multilayer stacking order of graphite. Unfortunately, superconductivity has not been experimentally observed in such systems. So the results in this section might not be as experimentally relevant for the time being. Nonetheless, from a theory point of view, these systems also host robust Majorana zero modes due to mirror symmetries if they have $f$-wave superconductivity.

\subsection{$ABA$ trilayer graphene}

\begin{figure}
    \centering
    \includegraphics[width=7in]{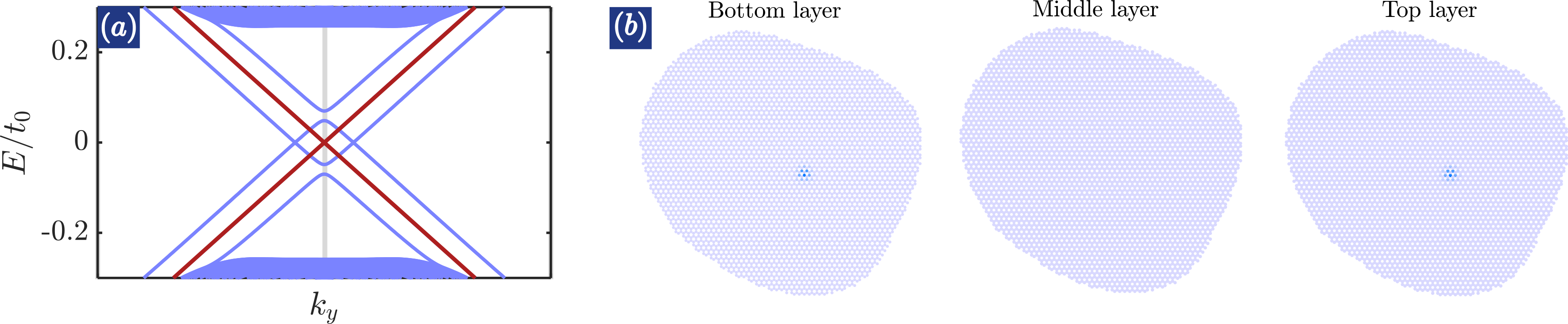}
    \caption{\textbf{$ABA$ trilayer graphene. } (a) Band structure of an armchair nanoribbon where the red bands are topological and come from the monolayer sector of the Hamiltonian. (b) A zero mode confined to a vortex with energy $E \approx 10^{-16}t_0$. The superconducting gap here is quite large $\Delta = 0.3t_0$ to facilitate fast computation.}
    \label{fig:ABAtrilayer}
\end{figure}

The Hamiltonian for $ABA$ trilayer graphene looks almost exactly like that for rhombohedral trilayer graphene with one important exception in the interlayer hopping
\begin{equation}
    \mathcal{H}^\text{BdG}(\mathbf{k}) = \begin{pmatrix}
    \mathcal{H}_3 (\mathbf{k}) & \mathcal{H}_\text{sc}(\mathbf{k}) \\
    \mathcal{H}_\text{sc}^\dagger(\mathbf{k}) & -\mathcal{H}_3 (\mathbf{k})
    \end{pmatrix},
\end{equation}
where 
\begin{equation}
    \begin{split}
        \mathcal{H}_3(\mathbf{k}) &= \begin{pmatrix}1 & 0 & 0\\
        0 & 1 & 0 \\
        0 & 0 & 1
        \end{pmatrix} \otimes \mathcal{H}_\text{intra}(\mathbf{k})  + \begin{pmatrix}0 & 1 & 0\\
        0 & 0 & 0\\
        0 & 1 & 0
        \end{pmatrix} \otimes \mathcal{H}_\text{inter}(\mathbf{k})  + \begin{pmatrix}0 & 0 & 0\\
        1 & 0 & 1\\
        0 & 0 & 0
        \end{pmatrix} \otimes \mathcal{H}^\dagger_\text{inter}(\mathbf{k}),  \\
        \mathcal{H}_\text{intra}(\mathbf{k}) &= \begin{pmatrix}
        \mu & -t_0 h_\mathbf{k} \\
        -t_0 h^*_\mathbf{k} & \mu
        \end{pmatrix}, \\
        \mathcal{H}_\text{inter}(\mathbf{k}) &= \begin{pmatrix}
        0 & 0 \\
        \gamma_1 & 0
        \end{pmatrix}, \\
        \mathcal{H}_\text{sc}(\mathbf{k}) &= \begin{pmatrix}1 & 0 & 0\\
        0 & 1 & 0 \\
        0 & 0 & 1
        \end{pmatrix} \otimes \begin{pmatrix}\Delta g_\mathbf{k} & 0\\
        0 & \Delta g_\mathbf{k}  
        \end{pmatrix}.
    \end{split} 
\end{equation}
This Hamiltonian importantly respects a mirror symmetry in the $z$ direction represented by 
\begin{equation}
    \mathcal{M}_z = \tau_0 \otimes \begin{pmatrix}
         0 & 0 & 1 \\
         0 & 1 & 0 \\
         1 & 0 & 0
    \end{pmatrix} \otimes \sigma_0.
\end{equation}
We have $\left[ \mathcal{H}^\text{BdG}(\mathbf{k}), \mathcal{M}_z \right] = 0.$ This symmetry is diagonal in $\mathbf{k}.$ So it holds for any $\mathbf{k}.$ Thus, we can block-diagonalize to write the Hamiltonian in the basis of eigenstates of $\mathcal{M}_z.$ In this new basis, the Hamiltonian 
\begin{equation}
     \mathcal{H}_3(\mathbf{k}) = \begin{pmatrix}1 & 0 & 0\\
        0 & 1 & 0 \\
        0 & 0 & 1
        \end{pmatrix} \otimes \mathcal{H}_\text{intra}(\mathbf{k})  + \begin{pmatrix}0 & 0 & 0\\
        0 & 0 & \sqrt{2}\\
        0 & 0 & 0
        \end{pmatrix} \otimes \mathcal{H}_\text{inter}(\mathbf{k})  + \begin{pmatrix}0 & 0 & 0\\
        0 & 0 & 0\\
        0 & \sqrt{2} & 0
        \end{pmatrix} \otimes \mathcal{H}^\dagger_\text{inter}(\mathbf{k}).
\end{equation}
This decomposition is similar to that done in Ref. \cite{khalaf2019magic}. As one can see, in this new basis, the system looks like one decoupled effective monolayer system and one decoupled effective bilayer graphene. The monolayer ($\mathcal{M}_z$-odd) sector is invariant under $\mathcal{M}_y.$ We note that the whole Hilbert space is \textit{not} invariant under $\mathcal{M}_y,$ only the mirror-odd sector is. So we can again classify the monolayer bands by the mirror-projected winding number as before. This predicts that everything we have learned in the monolayer model still applies to the mirror-odd sector of $ABA$ trilayer graphene. In particular, in armchair nanoribbons, we find a pair of topological zero modes at $k_y = 0$ as shown in Fig. \ref{fig:ABAtrilayer}(a). We also find two zero modes confined to a vortex, as shown in Fig. \ref{fig:ABAtrilayer}(b). We note that the zero mode is localized only in the top and bottom layers because it comes from the monolayer sector, which in this case, is just an odd combination of the top and bottom layers.

\subsection{$ABA$ odd-layer graphene}

\begin{figure}
    \centering
    \includegraphics[width=6in]{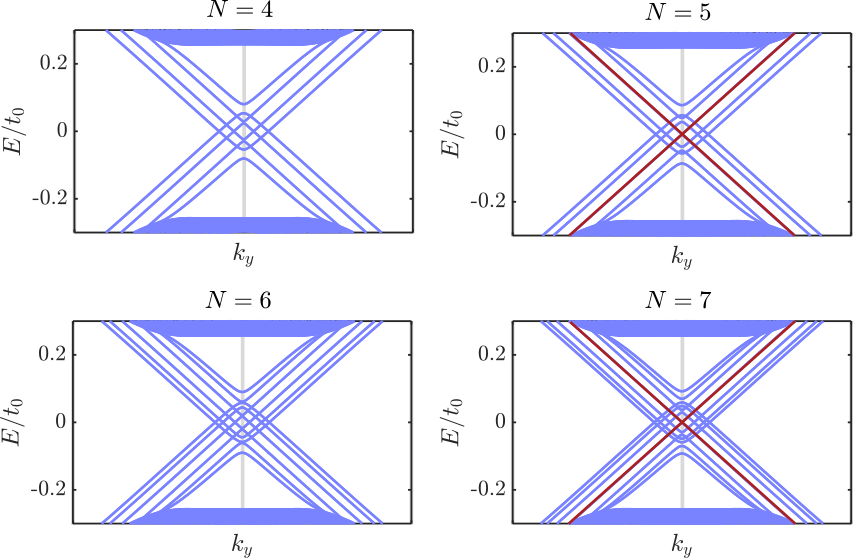}
    \caption{\textbf{Band structure of $ABA$ multilayer graphene.} We see that for $N$ odd, there are two modes crossing  $k_y = 0$ at exactly zero energy. Here, $\mu = \Delta = 0.1t_0.$}
    \label{fig:ABAmultilayer}
\end{figure}

We generalize the previous result to $ABA$ odd-layer graphene. The Hamiltonian is 

\begin{equation}
    \mathcal{H}^\text{BdG}(\mathbf{k}) = \begin{pmatrix}
    \mathcal{H}_N (\mathbf{k}) & \mathcal{H}_\text{sc}(\mathbf{k}) \\
    \mathcal{H}_\text{sc}^\dagger(\mathbf{k}) & -\mathcal{H}_N (\mathbf{k})
    \end{pmatrix},
\end{equation}
where 
\begin{equation}
    \begin{split}
        \mathcal{H}_N(\mathbf{k}) &= \mathbb{I}_{N\times N} \otimes \mathcal{H}_\text{intra}(\mathbf{k})  + \begin{pmatrix} 0 & 1 & 0 & 0 & 0 & 0 & 0 & \dots \\
                                        0 & 0 & 0 & 0 & 0 & 0 & 0 & \dots \\
                                        0 & 1 & 0 & 1 & 0 & 0 & 0 & \dots \\
                                        0 & 0 & 0 & 0 & 0 & 0 & 0 & \dots \\
                                        0 & 0 & 0 & 1 & 0 & 1 & 0 & \dots \\
                                        0 & 0 & 0 & 0 & 0 & 0 & 0 & \dots \\
                                        0 & 0 & 0 & 0 & 0 & 1 & 0 & \dots \\
        \vdots & \vdots & \vdots & \vdots & \vdots & \vdots & \vdots & \ddots \\
        \end{pmatrix} \otimes \mathcal{H}_\text{inter}(\mathbf{k})  + \begin{pmatrix} 0 & 0 & 0 & 0 & 0 & 0 & 0 & \dots \\
                                        1 & 0 & 1 & 0 & 0 & 0 & 0 & \dots \\
                                        0 & 0 & 0 & 0 & 0 & 0 & 0 & \dots \\
                                        0 & 0 & 1 & 0 & 1 & 0 & 0 & \dots \\
                                        0 & 0 & 0 & 0 & 0 & 0 & 0 & \dots \\
                                        0 & 0 & 0 & 0 & 1 & 0 & 1 & \dots \\
                                        0 & 0 & 0 & 0 & 0 & 0 & 0 & \dots \\
        \vdots & \vdots & \vdots & \vdots & \vdots & \vdots & \vdots & \ddots \\
        \end{pmatrix} \otimes \mathcal{H}^\dagger_\text{inter}(\mathbf{k}),  \\
        \mathcal{H}_\text{intra}(\mathbf{k}) &= \begin{pmatrix}
        \mu & -t_0 h_\mathbf{k} \\
        -t_0 h^*_\mathbf{k} & \mu
        \end{pmatrix}, \\
        \mathcal{H}_\text{inter}(\mathbf{k}) &= \begin{pmatrix}
        0 & 0 \\
        \gamma_1 & 0
        \end{pmatrix}, \\
        \mathcal{H}_\text{sc}(\mathbf{k}) &= \mathbb{I}_{N\times N} \otimes \begin{pmatrix}\Delta g_\mathbf{k} & 0\\
        0 & \Delta g_\mathbf{k}  
        \end{pmatrix}.
    \end{split} 
\end{equation}
We assume that only nearest-layer hoppings and intralayer hoppings are allowed. The rest of the discussion depends on this physically-motivated assumption. This Hamiltonian is invariant under $\mathcal{M}_z$
\begin{equation}
    \mathcal{M}_z = \tau_0 \otimes \bar{\mathbb{I}}_{N\times N} \otimes \sigma_0,
\end{equation}
where $\bar{\mathbb{I}}_{N\times N}$ is the \textit{anti}-diagonal $N\times N$ matrix. There is one unique state 
\begin{equation}
    \ket{\psi_\lambda} = \sqrt{\frac{2}{N+1}}\begin{pmatrix}
        1 & 0 & -1 & 0 & 1 & 0 & -1 & 0 & ...
    \end{pmatrix}^T
\end{equation}
that does \textit{not} couple to the interlayer hoppings. Explicitly, we observe 
\begin{equation}
    \begin{pmatrix} 0 & 1 & 0 & 0 & 0 & 0 & 0 & \dots \\
                                        0 & 0 & 0 & 0 & 0 & 0 & 0 & \dots \\
                                        0 & 1 & 0 & 1 & 0 & 0 & 0 & \dots \\
                                        0 & 0 & 0 & 0 & 0 & 0 & 0 & \dots \\
                                        0 & 0 & 0 & 1 & 0 & 1 & 0 & \dots \\
                                        0 & 0 & 0 & 0 & 0 & 0 & 0 & \dots \\
                                        0 & 0 & 0 & 0 & 0 & 1 & 0 & \dots \\
        \vdots & \vdots & \vdots & \vdots & \vdots & \vdots & \vdots & \ddots \\
        \end{pmatrix}\begin{pmatrix}
        1 \\
        0 \\
        -1 \\
        0 \\
        1 \\
        0 \\ 
        -1\\
        \vdots
    \end{pmatrix} = \begin{pmatrix} 0 & 0 & 0 & 0 & 0 & 0 & 0 & \dots \\
                                        1 & 0 & 1 & 0 & 0 & 0 & 0 & \dots \\
                                        0 & 0 & 0 & 0 & 0 & 0 & 0 & \dots \\
                                        0 & 0 & 1 & 0 & 1 & 0 & 0 & \dots \\
                                        0 & 0 & 0 & 0 & 0 & 0 & 0 & \dots \\
                                        0 & 0 & 0 & 0 & 1 & 0 & 1 & \dots \\
                                        0 & 0 & 0 & 0 & 0 & 0 & 0 & \dots \\
        \vdots & \vdots & \vdots & \vdots & \vdots & \vdots & \vdots & \ddots \\
        \end{pmatrix}\begin{pmatrix}
        1 \\
        0 \\
        -1 \\
        0 \\
        1 \\
        0 \\ 
        -1\\
        \vdots
    \end{pmatrix} = \mathbf{0}.
\end{equation}
Notice that this is only true if $N$ is odd. Let us call these two sparse matrices $A$ and $B.$ Then, we can write the full wavefunction as $\ket{\psi} = \ket{\psi_\text{monolayer}(\mathbf{k})}\otimes \ket{\psi_\lambda},$ where $\ket{\psi_\text{monolayer}(\mathbf{k})}$ is the wavefunction in a single layer, and observe that, temporarily switching the order of the Kronecker products,
\begin{equation}
\begin{split}
    &\left[ \tau_z \otimes \mathcal{H}_\text{intra}(\mathbf{k}) \otimes  \mathbb{I}_{N\times N} + \tau_x \otimes \sigma_0 \Delta g_\mathbf{k} \otimes \mathbb{I}_{N\times N} + \tau_z \otimes \mathcal{H}_\text{inter}(\mathbf{k}) \otimes A  + \tau_z \otimes \mathcal{H}^\dagger_\text{inter}(\mathbf{k}) \otimes B \right] \ket{\psi_\text{monolayer}(\mathbf{k})}\otimes \ket{\psi_\lambda} \\
    &=\left[ \tau_z \otimes \mathcal{H}_\text{intra}(\mathbf{k}) \otimes  \mathbb{I}_{N\times N} + \tau_x \otimes \sigma_0 \Delta g_\mathbf{k} \otimes \mathbb{I}_{N\times N} \right] \ket{\psi_\text{monolayer}(\mathbf{k})}\otimes \ket{\psi_\lambda} \\
    &= \left(\left[ \tau_z \otimes \mathcal{H}_\text{intra}(\mathbf{k})  + \tau_x \otimes \sigma_0 \Delta g_\mathbf{k}  \right] \ket{\psi_\text{monolayer}(\mathbf{k})} \right)\otimes \ket{\psi_\lambda}.
\end{split}
\end{equation}
Thus, if $\ket{\psi_\text{monolayer}(\mathbf{k})}$ is an eigenstate of $\tau_z \otimes \mathcal{H}_\text{intra}(\mathbf{k})  + \tau_x \otimes \sigma_0 \Delta g_\mathbf{k},$ then $\ket{\psi}$ is an eigenstate of the full Hamiltonian. We refer to states in this basis as coming from the effective monolayer sector because the Hamiltonian governing these states looks exactly like the monolayer Hamiltonian. This sector hosts topological zero modes at armchair edges, as shown in Fig. \ref{fig:ABAmultilayer}. We see from there that these modes do not exist in even-layer configurations, which is consistent with this analysis.

\section{Twisted multilayer graphene}

\subsection{Additional results for twisted bilayer graphene}
\begin{figure}
    \centering
    \includegraphics[width=5.6in]{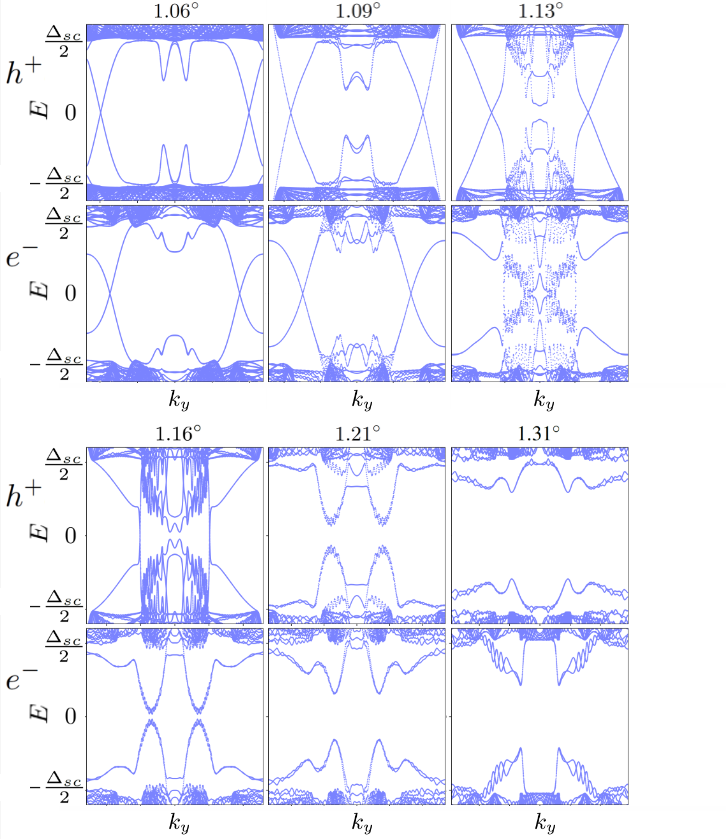}
    \caption{\textbf{Band structures of twisted bilayer graphene nanoribbons at different twist angles.} The top rows correspond to superconductivity in the hole dome (the gap opens at filling $n=-2.4$), and the bottom rows correspond to the electron dome (gap at filling $n=2.4$).\ In all cases, $\Delta_\text{sc}=1$ meV and the the width of the ribbon is $W\approx35 L_\text{M}$.}
    \label{fig:TBG SM}
\end{figure}

Here, we present band structures for twisted bilayer graphene (TBG) nanoribbons at six different twist angles and for both the hole and electron superconducting domes. As shown in Fig. \ref{fig:TBG SM}, the sub-gap spectra depend very sensitively on twist angle. Starting at the magic angle (1.06$^{\circ}$ in the model), increments as small as $0.1-0.2^{\circ}$ eliminate the zero modes and deplete the energy window around the center of the gap. As pointed out in the main text, the overall dynamic is the following: near the magic angle there are 8 zero modes; as the angle is increased, the momenta of these modes diminish; increasing the angle further eliminates these modes; and at the largest angle of 1.31$^{\circ}$, all Andreev states with energy $|\epsilon|<\Delta_\text{sc}/4$ have disappeared. This evolution is very similar in the hole and electron domes (compare top and bottom rows in Fig. \ref{fig:TBG SM}). However, there is electron-hole asymmetry, which manifests quantitatively at every angle and qualitatively in that the zero modes in the hole dome are more robust, in the sense that they disappear at a larger twist angle.

\subsection{Details of the model}
The TBG nanoribbon unit cell is built following the procedure of Ref. \cite{sainzcruz21high}. In the notation of that paper, we build a TBG nanoribbon with chiral vectors $(440,20)@(-440,-20)$, which has a twist angle of 4.41$^{\circ}$, a width of $W\approx35 L_\text{M}$, and $27,040$ sites in its unit cell. We have checked that the band structures have converged with respect to ribbon width. One detail to take into account is that in the construction of Ref. \cite{sainzcruz21high} the origin of coordinates is at an atom, while in the present case it is necessary to set the origin at the center of a graphene hexagon, for the resulting system to have $\mathcal{M}_y$ symmetry.

\begin{figure}
    \centering
    \includegraphics[scale=0.7]{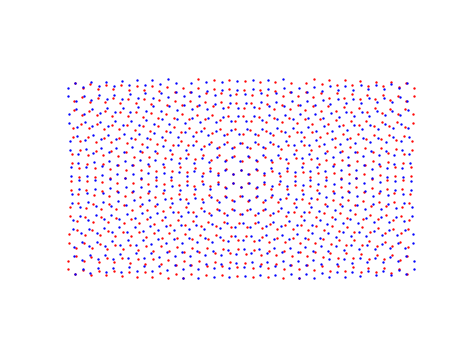}
    \caption{\textbf{Unit cell of a TBG nanoribbon with chiral vector $(22,1)@(-22,-1)$}. Joining 20 such cells horizontally gives the unit cell for the nanoribbon with chiral vector $(440,20)@(-440,-20)$, of width $W\approx35 L_\text{M}$, for which we obtain the spectrum in Fig. 3(a, c) in the main text and Fig. \ref{fig:TBG SM} above.}
    \label{fig:TBGedge}
\end{figure}

We start from the Lin-Tománek tight-binding, non-interacting Hamiltonian ${\cal H}_{\theta}$ \cite{lin2018minimum} and, for TBG, we include Hartree electron-electron interactions through an electrostatic potential~\cite{guinea18electrostatic}. Then, we build $\mathcal{H}^\text{BdG}$ with the same superconducting hoppings as before:
\begin{equation}
{{\cal H}_{\theta}=-\sum_{i\neq j,m}\gamma_{ij}^{mm}(\hat{c}^{\dagger}_{i,m}\hat{c}_{j,m}+\text{h.c.})}-\sum_{i, j,m}\gamma_{ij}^{m,m+1}(\hat{c}^{\dagger}_{i,m}\hat{c}_{j,m+1}+\text{h.c.})+\sum_{i,m}V_\text{H}(n)\hat{c}^{\dagger}_{i,m}\hat{c}_{i,m}\, ,
\label{eq:s3}
\end{equation}
where $i,j$ run over the lattice sites and $m$ is the layer index. ${\cal{H}}_{\theta}$ includes intralayer hopping to nearest-neighbors only $\gamma_{ij}^{mm}=t_0$ and interlayer hopping that decays exponentially away from the vertical direction, $\gamma_{ij}^{m,m+1}=t_{\perp}e^{-(\sqrt{r^2+d^2}-d)/\lambda_\perp}\frac{d^2}{r^2+d^2}$, where $d=0.335$ nm is the distance between layers, $t_0=3.09$ eV and $t_{\perp}=0.39$ eV are the intralayer and interlayer hopping amplitudes, and $\lambda_\perp=0.027$ nm is a cutoff for the interlayer hopping \cite{lin2018minimum}.

The parameters in the tight binding model are scaled, so that the central bands of TBG with twist angle $\theta$ are approximated by the central bands of an equivalent lattice with twist angle $\lambda \theta$, with $\lambda >1$~\cite{gonzalez2017,vahedi2021magnetism,sainzcruz21high,sainzcruz2023junctions}. This scaling approximation is based on the fact that the Dirac equation that governs each layer can be described in different ways, and one of them is a honeycomb lattice of super-atoms, each representing a small cluster of atoms of the original lattice. In this scaled system, the lattice constant is magnified and the hopping reduced. In the context of twisted bilayer graphene, the possibility of scaling manifests also in the continuum model, which shows that the bands depend, to first order, on a dimensionless parameter \cite{bistritzer2011moire},
\begin{equation}
\alpha = \frac{at_{\perp}}{2\hbar v_F \sin ( \theta / 2 )} \propto \frac{t_{\perp}}{t_0\theta}\, .
\label{eq:s4}
\end{equation}
where $a$ is the lattice constant and $v_F$ is the Fermi velocity. This parameter can be understood as a comparison between the time a carrier needs to traverse a unit cell within a layer, and the average time between interlayer tunneling events.
Thus, a small angle $\theta$ can be simulated with a larger one $\theta^{\prime}$ by doing the following transformations: $t_0\rightarrow\frac{1}{\lambda} t_0$, $a\rightarrow\lambda a$, $d\rightarrow\lambda d$, with $\lambda=\sin(\frac{\theta^{\prime}}{2})/\sin(\frac{\theta}{2})$ \cite{gonzalez2017,vahedi2021magnetism,sainzcruz21high, sainzcruz2023junctions}. Note that the interlayer distance is scaled to keep the interlayer hopping unchanged after scaling.\ This approximation reproduces well the low-energy band structure, as shown in Refs. \cite{vahedi2021magnetism,sainzcruz2023junctions}.

In Eq. (\ref{eq:s3}), the Hartree term is 
\begin{equation} V_\text{H}(n)=\frac{2 \rho(n)}{\epsilon_r L_\text{M}}\sum^3_{i=1}\cos(\mathbf{G}_i \cdot \mathbf{r}),
\end{equation}
where $\mathbf{G}_i$ are the reciprocal lattice vectors, $\mathbf{r}$ the position, $L_\text{M}$ the moiré period, $\epsilon_r=4$ the dielectric constant due to hBN encapsulation, and $\rho(n)$ a filling dependent parameter, listed in Table \ref{table1}. To obtain $\rho(n)$, we fit the band structure of the tight-binding model to the continuum model of Ref. \cite{moon2014optical} and do a self-consistent calculation.

\begin{table}[h]
\begin{tabular}{|p{1.5cm}||p{1cm}|p{1cm}|p{1cm}|p{1cm}|p{1cm}|}
 \hline
 \multicolumn{6}{|c|}{Values of $\rho$ for the Hartree term} \\
 \hline
 & $1.06^{\circ}$ & $1.09^{\circ}$& $1.13^{\circ}$& $1.21^{\circ}$& $1.31^{\circ}$\\
 \hline
 $n=-2.4$ & $-0.768$ & $-0.776$ & $-0.773$ & $-0.742$ & $-0.601$\\
 \hline
 $n=+2.4$ & $0.768$ & $0.795$ & $0.793$ & $0.756$ & $0.612$\\
 \hline
\end{tabular}
\caption{Calculated values of the constant $\rho$ as a function of twist angle, filling and dielectric constant.}
\label{table1}
\end{table}

The reason why we include the Hartree term only in TBG, but not in twisted trilayer graphene (TTG), is that in TBG, we aim at a simulation with realistic parameters, in particular with a superconducting gap of $\Delta_\text{sc}=1$ meV, so it is worthwhile to include the Hartree interactions, which are a relevant correction to the non-interacting band structure.

\begin{figure}
    \centering
    \includegraphics[width=2.6in]{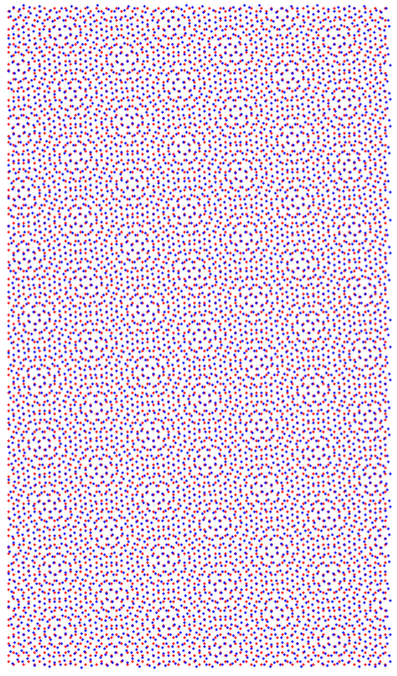}
    \caption{\textbf{Unit cell of a TTG nanoribbon with chiral vector $(-37,0)@(33,7)@(-37,0)$}. The translational vector of the middle layer is commensurate (exactly 37 times larger) with the translational vector of the top and bottom armchair layers. Joining 5 such cells horizontally gives the unit cell for the TTG nanoribbon for which we obtain the spectrum in Fig. 3(b) in the main text.}
    \label{fig:TTGcommesuration}
\end{figure}

In contrast, in the TTG nanoribbon we discuss, cutting armchair edges in the top and bottom layers, and a `chiral' edge on the middle layer, leads to a stringent commensurability condition: the translational modulus of the chiral layer needs to be a multiple of the translational modulus of the armchair layers, so that the overall translational symmetry is preserved. The translational modulus of a nanoribbon with chiral vector $\mathbf{C}=n\mathbf{a}_1+m\mathbf{a}_2$, with $\mathbf{a}_{1,2}=(\sqrt{3}/2,\pm1/2)$, is given by
\begin{equation}
    T=\frac{3}{4}(t_1+t_2)^2+\frac{1}{4}(t_1-t_2)^2
\end{equation}
with $t_1=(2m+n)/N_C$, $t_2=(2n+m)/N_C$, where $N_C=\text{gcd}\left(2m+n,2n+m\right)$ and gcd stands for the greatest common divisor. Therefore, the commensurability of the translational modulus of a ribbon with a generic chiral vector $(n,m)$ and an armchair ribbon with chiral vector $(n,0)$, is only fulfilled for some pairs $(n,m)@(n,0)$. One such pair is $(-37,0)@(33,7)$, for which $T_{(33,7)}=37T_{(-37,0)}$. As a consequence, the unit cell of the resulting TTG ribbon is large, see Fig. \ref{fig:TTGcommesuration}. This precludes a simulation of the full sub-gap spectrum with a realistic superconducting gap of $\Delta_\text{sc}=1$ meV, which would also require the ribbon to be wide enough not to mix the zero modes of the left and right edges, resulting in a system with many moiré unit cells.

Therefore, the TTG result is meant as a proof of concept: it shows that the decomposition of TTG into TBG plus monolayer graphene works for our model. This is evident from the fact that TTG features the four zero modes at $k_y=0$ found in monolayer graphene, and that these modes indeed come the effective monolayer that results from the odd combination of top and bottom layers \cite{khalaf2019magic}, as manifest in their charge distributions, which are identical in the top and bottom layers, and zero in the middle layer, as shown in Fig. \ref{fig:TTGribbonMZM} below. Moreover, as we have emphasized, the zero modes confined to vortex cores in twisted trilayer graphene are much more experimentally relevant, because edge engineering is very challenging, whereas vortices occur naturally in the superconductor and can be probed with STM measurements.

\begin{figure}
    \centering
    \includegraphics[width=2.7in]{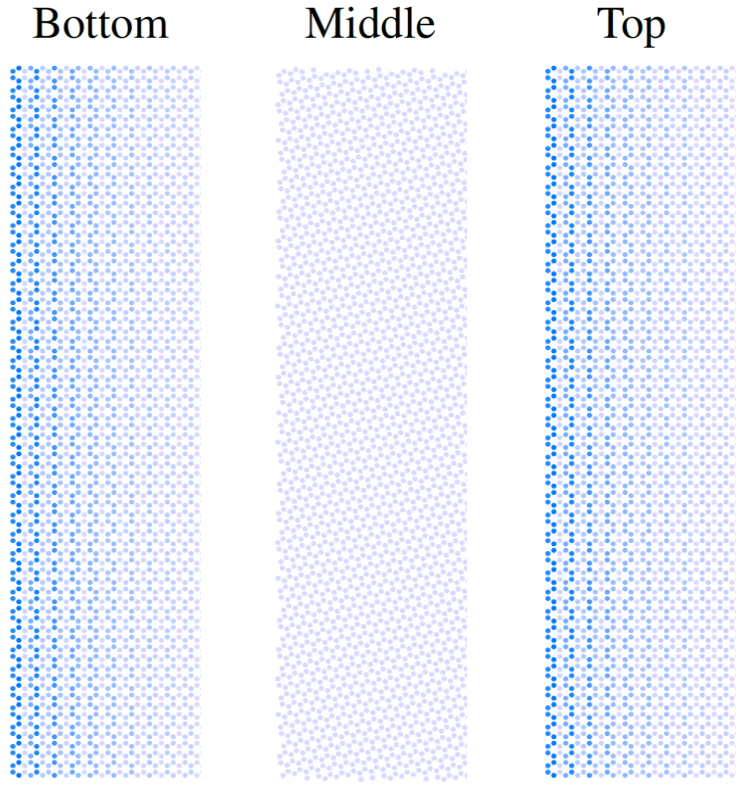}
    \caption{\textbf{Charge map of a zero mode in the TTG nanoribbon.} The charge is plotted in each of the three layers, near the left edge of the ribbon. This is one of the four zero modes at $k_y=0$ in Fig. 3(b) in the main text.}
    \label{fig:TTGribbonMZM}
\end{figure}

\bibliography{References.bib}

\begin{thebibliography}{95}
\bibitem [{TTG()}]{TTGparameters}%
  \BibitemOpen
  \href@noop {} {}\bibinfo {note} {This twist angle and superconducting gap are
  not realistic for TTG and are chosen only to facilitate the convergence of
  the computation.}\BibitemShut {Stop}%
\end{thebibliography}

\end{document}